\documentstyle[epsfig,a4,12pt,rotating,here,amssymb]{article}
\def\mrm{\mathrm}

\newcommand{\eerhz}{\mbox{$\mathrm{e}^+\mathrm{e}^-\rightarrow\mathrm{h}^{0}\mathrm{Z}^{0}$}}

\newcommand{\glgl}{\mbox{$\mathrm{g}\mathrm{g}$}}

\newcommand{\Z}{\mbox{$\mathrm{Z}^{0}$}}
\newcommand{\A}{\mbox{$\mathrm{A}^{0}$}}

\newcommand{\Hosm}{\mbox{$\mathrm{H}^{0}_{\mathrm{SM}}$}}
\newcommand{\Wp}{\mbox{$\mathrm{W}^+$}}
\newcommand{\Wm}{\mbox{$\mathrm{W}^-$}}
\newcommand{\Hpm}{\mbox{$\mathrm{H}^{\pm}$}}

\newcommand{\ZZ}{\mbox{$\mathrm{Z}^{0}{\mathrm{Z}^{0}}^{(*)}$}}

\newcommand{\cc}{\mbox{$\mathrm{c} \bar{\mathrm{c}}$}}

\newcommand{\mWn}{\mbox{$m_{\mathrm{W}}$}}

\newcommand{\G}{\mbox{$\mathrm{GeV}$}}

\newcommand{\ie}{\mbox{$i.e.$}}
\newcommand{\sba}{\mbox{$\sin ^2 (\beta -\alpha)$}}
\newcommand{\cba}{\mbox{$\cos ^2 (\beta -\alpha)$}}

\newcommand{\ee}{\mbox{${\mathrm{e}}^+ {\mathrm{e}}^-$}}
\newcommand{\tautau}{\mbox{$\tau^+\tau^-$}}
\newcommand{\mm}{\mbox{$\mu^+\mu^-$}}

\newcommand{\qq}         {\mbox{$\mathrm{q}\bar{\mathrm{q}}$}}
\newcommand{\bb}         {\mbox{$\mathrm{b}\bar{\mathrm{b}}$}}

\newcommand{\mZ}         {\mbox{$m_{\mathrm{Z}}$}}
\newcommand{\mH}         {\mbox{$m_{\mathrm{H}}$}}
\newcommand{\mh}         {\mbox{$m_{\mathrm{h}}$}}
\newcommand{\mA}         {\mbox{$m_{\mathrm{A}}$}}

\newcommand {\Ho}        {\mbox{$\mathrm{H}^{0}$}}
\newcommand {\Ao}        {\mbox{$\mathrm{A}^{0}$}}
\newcommand {\ho}        {\mbox{$\mathrm{h}^{0}$}}
\newcommand {\Zo}        {\mbox{$\mathrm{Z}^{0}$}}

\newcommand{\nn}{\mbox{$\nu \bar{\nu}$}}

\newcommand{\WW}         {\mbox{$\mathrm{W}^+\mathrm{W}^-$}}
\newcommand{\pb}         {\mbox{$\mathrm{pb}^{-1}$}}
\newcommand{\tanb}       {\mbox{$\tan\!\beta$}}
\newcommand{\h}{\mbox{$\mathrm{h}^{0}$}}
\def\mrm       {\mathrm}

\newcommand{\sqrts}     {\mbox{$\sqrt{s}$}}

\newcommand{\ra}        {\mbox{$\rightarrow$}}   

\newcommand{\me}{matrix element}

\begin{document}
\begin{titlepage}
\centerline{\Large EUROPEAN ORGANIZATION FOR NUCLEAR RESEARCH}
\bigskip
\begin{flushright}
      CERN-EP-2000-092\\ 
      July 7, 2000
\end{flushright}
\bigskip\bigskip\bigskip

\begin{center}{\Large\bf  Two Higgs Doublet Model and Model Independent \\
                          Interpretation of Neutral Higgs Boson Searches}
\end{center}
\bigskip
\begin{center}{\large The OPAL Collaboration}
\end{center}

\bigskip
\begin{center}{\large  Abstract}\end{center}

\hspace{-0.75cm}

Searches for the neutral Higgs bosons \ho\ and \Ao\,
are used to obtain limits on 
the Type II Two Higgs Doublet Model (2HDM(II)) with 
no CP--violation in the Higgs sector and no additional particles 
besides the five Higgs bosons.
The analysis combines approximately 
170 pb$^{-1}$ of data collected with the OPAL detector 
at $\sqrt{s} \approx 189$~GeV  with 
previous runs at $\sqrt{s} \approx m_{Z}$ and 
$\sqrt{s} \approx 183$ GeV.
The searches are sensitive to the
\ho, \Ao\ra\qq, \glgl, \tautau\ and \ho\ra\Ao\Ao\ decay modes 
of the Higgs bosons. 
For the first time, the 2HDM(II) parameter space is explored 
in a detailed scan, and
new flavour independent analyses are applied  
to examine regions in which 
the neutral Higgs bosons decay predominantly into 
light quarks or gluons. Model--independent limits are also given.

\bigskip\bigskip
\bigskip\bigskip\bigskip\bigskip
\bigskip\bigskip\bigskip\bigskip
\bigskip\bigskip\bigskip\bigskip
\bigskip\bigskip\bigskip\bigskip

\begin{center}{\large
(Submitted to European Physical Journal C)
}\end{center}
\end{titlepage}
\begin{center}{\Large        The OPAL Collaboration
}\end{center}\bigskip
\begin{center}{
G.\thinspace Abbiendi$^{  2}$,
K.\thinspace Ackerstaff$^{  8}$,
C.\thinspace Ainsley$^{  5}$,
P.F.\thinspace {\AA}kesson$^{  3}$,
G.\thinspace Alexander$^{ 22}$,
J.\thinspace Allison$^{ 16}$,
K.J.\thinspace Anderson$^{  9}$,
S.\thinspace Arcelli$^{ 17}$,
S.\thinspace Asai$^{ 23}$,
S.F.\thinspace Ashby$^{  1}$
% ...
}\end{center}\bigskip
\bigskip
$^{  1}$School of Physics and Astronomy, University of Birmingham,
Birmingham B15 2TT, UK
\newline
$^{  2}$Dipartimento di Fisica dell' Universit\`a di Bologna and INFN,
I-40126 Bologna, Italy
\newline
$^{  3}$Physikalisches Institut, Universit\"at Bonn,
D-53115 Bonn, Germany
\newline
% ...
$^{  5}$Cavendish Laboratory, Cambridge CB3 0HE, UK
\newline
% ...
$^{  8}$CERN, European Organisation for Nuclear Research,
CH-1211 Geneva 23, Switzerland
\newline
$^{  9}$Enrico Fermi Institute and Department of Physics,
University of Chicago, Chicago IL 60637, USA
\newline
% ...
$^{ 16}$Department of Physics, Schuster Laboratory, The University,
Manchester M13 9PL, UK
\newline
$^{ 17}$Department of Physics, University of Maryland,
College Park, MD 20742, USA
\newline
% ...
$^{ 22}$Department of Physics and Astronomy, Tel Aviv University,
Tel Aviv 69978, Israel
\newline
$^{ 23}$International Centre for Elementary Particle Physics and
Department of Physics, University of Tokyo, Tokyo 113-0033, and
Kobe University, Kobe 657-8501, Japan
\newline
% ...

\newpage
\section{Introduction}
\label{sect:intro}
In this study approximately 170~\pb\ of the data\footnote{The searches 
presented here
use subsets of the data sample for which the necessary
detector components were fully operational.  At $\sqrt{s} \approx$ 189
GeV approximately 188~\pb\ were
collected and 170 \pb\ analysed, varying by $\pm$2\% from channel to channel, 
depending on the detector components required.}
collected by the OPAL detector at LEP at 
$189$~GeV 
centre-of-mass energy are combined
with 58 \pb\ of data taken at 
the \Zo\ pole and 53 \pb\ of data at \sqrts~$\approx$ 
183 GeV to search for 
neutral Higgs bosons \cite{higgs, higgsEng, higgsGur} 
in the framework of the Type II Two Higgs Doublet Model  
with no CP--violation in the Higgs sector 
and no additional particles besides those arising from the Higgs mechanism
(2HDM(II)) \cite{hollik, hollik2}.
A model--independent scheme, in which 
no assumption is made on the structure of the Higgs sector,
is also analysed.

In the minimal Standard Model (SM) the Higgs sector 
comprises only one complex Higgs doublet \cite{higgs}
resulting in one physical neutral Higgs 
scalar whose mass is a free parameter of the theory. 
However, since there is no experimental evidence 
for the Higgs boson, 
it is important to study extended models  
containing more than one physical Higgs boson 
in the spectrum. 
In particular, Two Higgs Doublet Models (2HDMs) are attractive extensions of the SM
since they add new phenomena with 
the fewest new parameters; they satisfy the 
constraints of $\rho \approx 1$ and the absence 
of tree-level flavour changing neutral currents,
if the Higgs-fermion couplings are appropriately chosen.
In the context of 2HDMs
the Higgs sector comprises five physical Higgs bosons: 
two neutral CP-even scalars, \ho\ and \Ho\ (with  \mh\ $<$ \mH), one
CP-odd scalar, \Ao, and two charged scalars, \Hpm. 

The most general CP--invariant Higgs potential, having two complex $Y=1$, $SU(2)_L$ doublet 
scalar fields $\phi_1$ and $\phi_2$, is given by \cite{hollik, hollik2,higgshunter}
$${{V(\phi_1, \phi_2)}}={\kappa_1} (\phi_1^{\dagger}\phi_1 - { v_1^2})^2 
                             +{\kappa_2} (\phi_2^{\dagger}\phi_2 -{ v_2^2})^2  
                             +{\kappa_3} [(\phi_1^{\dagger}\phi_1 - { v_1^2}) +
                                 (\phi_2^{\dagger}\phi_2 - { v_2^2})]^2$$
\begin{equation}
       +{\kappa_4} [(\phi_1^{\dagger}\phi_1)(\phi_2^{\dagger}\phi_2) -
                                 (\phi_1^{\dagger}\phi_2)(\phi_2^{\dagger}\phi_1)] 
                      + {\kappa_5}[{\mathrm{Re}}(\phi_1^{\dagger}\phi_2) - v_1 v_2]^2 
                      + {\kappa_6} [{\mathrm{Im}}(\phi_1^{\dagger}\phi_2)]^2, 
\end{equation}
where the vacuum expectation values, $v_i$, are non--negative 
real parameters and the couplings, $\kappa_i$, are real parameters.
The physical masses at tree level are given by:
\begin{eqnarray}
m^2_{{\mathrm{H,h}}} 
= \frac{1}{2}[{\mathcal{M}}_{11} + {\mathcal{M}}_{22} \pm \sqrt{({\mathcal{M}}_{11} - {\mathcal{M}}_{22})^2 
+ 4 {\mathcal{M}}^2_{12}}], \\
m_{\mathrm{A}}^2 = \kappa_6 (v_1^2+ v_2^2), ~~~ m_{\mathrm{H}^\pm}^2 = \kappa_4 (v_1^2+ v_2^2),
\end{eqnarray}
where
\begin{eqnarray}
{\mathcal{M}}_{11} = 4(\kappa_1+\kappa_3)v_1^2+\kappa_5 v_2^2 \\
{\mathcal{M}}_{22} = 4(\kappa_2+\kappa_3)v_2^2+\kappa_5 v_1^2 \\
{\mathcal{M}}_{12} =  (4\kappa_3+\kappa_5)v_1 v_2. 
\end{eqnarray}
The Higgs mixing angle, $\alpha$,
is obtained from
\begin{eqnarray}
\cos{2\alpha} = \frac{{\mathcal{M}}_{11} - {\mathcal{M}}_{22}}{\sqrt{({\mathcal{M}}_{11} - {\mathcal{M}}_{22})^2 + 4 {\mathcal{M}}_{12}^2}}, \\
\sin{2\alpha} = \frac{2{\mathcal{M}}_{12}}{\sqrt{({\mathcal{M}}_{11} - {\mathcal{M}}_{22})^2 + 4 {\mathcal{M}}_{12}^2}},
\end{eqnarray}
and the angle $\beta$ is defined as the ratio of the
vacuum expectation values, $v_1$ and $v_2$, of the two scalar fields,
$\tanb=v_2/v_1$, with $0 \le \beta \le \pi/2$. 

At the centre-of-mass energies accessed by
LEP, the \ho\ and \Ao\  
bosons are expected to be produced predominantly via two processes: 
the {\it{Higgs--strahlung}}
process \ee\ra\ho\Zo\ 
and the {\it{pair--production}} process \ee\ra\ho\Ao.
The cross-sections for these two processes,
$\sigma_{\mathrm{hZ}}$ and $\sigma_{\mathrm{hA}}$,
are related at tree-level 
to the SM cross-sections by the following relations \cite{higgshunter}: 
\begin{eqnarray}
\ee\ra\ho\Zo\;:&&
\sigma_{\mathrm{hZ}}=\sin^2(\beta -\alpha)~\sigma^{\mathrm{SM}}_{\mathrm{HZ}},
\label{equation:xsec_zh} \\
\ee\ra\ho\Ao\;:&&
\sigma_{\mathrm{hA}}=
\cos^2(\beta-\alpha)~\bar{\lambda}~\sigma^{\mathrm{SM}}_{\mathrm{HZ}},
\label{equation:xsec_ah}
\end{eqnarray} 
where $\sigma^{\mathrm{SM}}_{\mathrm{HZ}}$ is the Higgs--strahlung cross-section 
for the SM process \ee\ra\Hosm\Zo, and
$\bar{\lambda} = \lambda^{3/2}_{\rm{Ah}} /\{{\lambda^{1/2}_{\rm{Zh}}
[12 m_{\rm{Z}}^2/s + \lambda_{\rm{Zh}}]}\} $
accounts for the suppression of the P-wave cross-section near the 
threshold, with 
$\lambda_{ij} = (1- m_i^2/s + m_j^2/s)^2 - 4 m_i^2m_j^2/s^2$
being the two--particle phase--space factor.

Within 2HDMs the choice of the couplings between the Higgs bosons and
the fermions determines the type of the model considered. In the Type II 
model the first Higgs doublet ($\phi_1$)
couples only to down--type fermions and the 
second Higgs doublet ($\phi_2$) couples only to up--type 
fermions. In the Type I model the 
quarks and leptons do not couple to the first Higgs 
doublet ($\phi_1$), but couple to the second Higgs 
doublet ($\phi_2$).
The Higgs sector in the 
minimal supersymmetric extension of the SM ~\cite{higgshunter,fayet}
is a Type II 2HDM, in which the introduction of 
supersymmetry adds new particles and constrains the parameter space of 
the model. 

In a 2HDM the production cross-sections and Higgs boson decay branching ratios
are predicted for a given set of model parameters.
The coefficients \sba\ and \cba\ which appear
in Eqs.~(\ref{equation:xsec_zh}) and~(\ref{equation:xsec_ah})
determine the production cross-sections.  The decay branching ratios to
the various final states 
are also determined by $\alpha$ and $\beta$.
In the 2HDM(II) the tree-level couplings of the \ho\ and \Ao\ bosons to the up-- and down--type 
quarks relative to the canonical SM values are 
\cite{higgshunter}
\begin{equation}
\label{eq:BRs}
{\mathrm{h^0}} {\mathrm{c}} \overline{{\mathrm{c}}} : \frac{\cos \alpha}{\sin\beta},~~~~
{\mathrm{h^0}} {\mathrm{b}} \overline{{\mathrm{b}}} : -~\frac{\sin \alpha}{\cos\beta},~~~~
{\mathrm{A^0}} {\mathrm{c}} \overline{{\mathrm{c}}} : \cot \beta,~~~~
{\mathrm{A^0}} {\mathrm{b}} \overline{{\mathrm{b}}} : \tan \beta,
\end{equation}
indicating the need for a scan over the range of both angles 
when considering the different production 
cross-section mechanisms and final state topologies.

In the analysis described in this paper, 
detailed scans over broad ranges of these parameters are performed.
Each of the scanned points is considered as an independent scenario
within the 2HDM(II), and results are provided for each point in the
(\mh,~\mA,~\tanb,~$\alpha$) space.
The final-state topologies of the processes (\ref{equation:xsec_zh}) and
(\ref{equation:xsec_ah}) are determined by the decays of the \Z,
\h\ and \A\ bosons. Higgs bosons couple to fermions 
with a strength proportional to the fermion mass, 
favouring the decays into pairs of 
b--quarks and tau leptons at LEP energies. 
However, with values of $\alpha$ and \tanb\ close to zero
the decays into up--type light quarks and gluons through quark loops become dominant,
motivating the development of new flavour independent analyses.

Section \ref{sect:detector} contains 
a short description of the OPAL detector and the Monte Carlo simulations used. 
The data samples and the final topologies studied
are discussed in Section \ref{sect:datasamples}.
The new flavour independent searches for \ee\ra\h\Z\ and 
\ee\ra\h\A\ are covered in Sections \ref{sect:zhsearches} and
\ref{sect:ahsearches}, respectively.
The model-independent and 2HDM interpretations of the searches 
are presented in Sections \ref{section:modindep} and 
\ref{section:limits}, respectively.
In Section \ref{section:conclusion} the results are summarised 
and conclusions are drawn. 

\section{OPAL detector and Monte Carlo samples}\label{sect:detector}
The OPAL detector~\cite{detector} has
nearly complete solid angle coverage and excellent hermeticity.
The innermost detector of the central tracking is a high-resolution
silicon microstrip vertex detector~\cite{simvtx} which lies immediately
outside of the beam pipe.  Its coverage in
polar angle\footnote{
OPAL uses a right-handed
coordinate system where the $+z$ direction is along the electron beam and
where $+x$ points to the centre of the LEP ring.  
The polar angle, $\theta$, is
defined with respect to the $+z$ direction and the azimuthal angle, $\phi$,
with respect to the horizontal, $+x$ direction.}
is $|\cos\theta|<0.9$.   The silicon microvertex detector
is surrounded by a high precision 
vertex drift chamber,
a large volume jet chamber, and $z$--chambers to measure the $z$ coordinates
of tracks, all in a uniform
0.435~T axial magnetic field. The lead-glass electromagnetic calorimeter
and the presampler are located outside the magnet coil.  It provides,
in combination with
the forward calorimeter, the gamma catcher, the MIP plug,
and the silicon-tungsten luminometer~\cite{sw}, geometrical acceptance
down to 25~mrad from the beam direction.  The silicon-tungsten luminometer
serves to measure the integrated luminosity using small angle Bhabha
scattering events~\cite{lumino}.
The magnet return yoke is instrumented with streamer tubes and thin gap
chambers for hadron calorimetry and is surrounded by several layers 
of muon chambers.

Events are reconstructed from charged particle tracks and
energy deposits (``clusters") in the electromagnetic and hadron calorimeters.
The tracks and clusters must pass a set of quality requirements
similar to those used in
previous OPAL Higgs boson searches~\cite{higgsold}.
In calculating the total visible energies and momenta, $E_{\mathrm vis}$
and $\vec{P}_{\mathrm vis}$, of events and
individual jets~\cite{drm}, corrections are applied to prevent the 
double counting of energy of tracks with associated
clusters~\cite{lep2neutralino}. 

A variety of Monte Carlo samples has been generated in order to estimate the
detection efficiencies for Higgs boson production and background from SM 
processes.
Higgs production is modelled with the HZHA generator~\cite{hzha}
for a wide range of Higgs masses.  The size of these samples
varies from 500 to 10,000 events.
The background processes are simulated, typically with more than 50 times the
statistics of the collected data,
by the following event generators:
PYTHIA~\cite{pythia} (\qq($\gamma$)), 
grc4f~\cite{grc4f} and for the study of the systematic errors EXCALIBUR~\cite{excalibur} 
(4-fermion processes);
BHWIDE~\cite{bhwide} (\ee$(\gamma)$);
KORALZ~\cite{koralz} (\mm$(\gamma)$ and \tautau$(\gamma)$);
PHOJET~\cite{phojet}; HERWIG~\cite{herwig}, and
Vermaseren~\cite{vermaseren} (hadronic and leptonic two-photon processes
($\gamma\gamma$)).
The hadronisation process is simulated with 
JETSET~\cite{pythia} with parameters
described in~\cite{opaltune}.
The cluster fragmentation model in HERWIG is used
to study the uncertainties due to quark jet fragmentation.
For each Monte Carlo sample, the detector response to the generated
particles is simulated in full detail~\cite{gopal}.

\section{Data samples and final state topologies studied}
\label{sect:datasamples}

The present study relies on the data collected by 
OPAL at  \sqrts~$\approx$ \mZ, 183 and 189 GeV.
The data collected at the \Z\ pole provide useful information in 2HDM
scenarios where the Higgs bosons are light; these data
have been extensively analysed  
in previous OPAL publications ~\cite{Z01, Z02, Z03}.
Higgs search results assuming SM decays from OPAL can be found 
in ~\cite{pn183} and ~\cite{pr285}
for \sqrts~$\approx$ 183 and 189 GeV, respectively.
In addition, at \sqrts~$\approx$  189 GeV, new flavour independent
channels are analysed for the first time to explore final 
state topologies in which
no assumption \thinspace is made \thinspace on the quark \thinspace flavours arising from
the
\begin{table}[H]
\vspace*{-0.4cm}
\begin{center}
\begin{tabular}{|c|c|c|c|c|} \hline
Channel \ho\Zo\ra & Luminosity [pb$^{-1}$] & Data & Total bkg. & Efficiency [\%]
 \\\hline\hline
\multicolumn{5}{|c|}{\bf\boldmath $\sqrts=183$~GeV} \\\hline\hline
{\small{\bb\qq}}  & 54.1     &7  &   $4.9 \pm 0.2 \pm 0.6$   &  $39.2 \pm 0.72 \pm 1.2$ \\ \hline
{\small{\bb\nn}}  &53.9 & 0 & $1.56 \pm 0.13 \pm 0.18$   &  $47.9 \pm 0.4 \pm 0.2$\\\hline
{\small{\bb\tautau}} &  &  &  & $41.7 \pm 2.5 \pm 1.8$
\\ \cline{1-1}\cline{5-5} 
{\small{\tautau\qq}} & \raisebox{1.5ex}[0cm][0cm]{53.7}     & \raisebox{1.5ex}[0cm][0cm]{1} & \raisebox{1.5ex}[0cm][0cm]{$ 1.3\pm 0.1 \pm 0.2$}      & $34.2 \pm 2.1 \pm 1.4$ \\\hline 
{\small{\bb\ee}}  &53.7  & 0 & $0.37 \pm 0.07 \pm 0.2 $  & $68.9 \pm 0.8 \pm 0.9$ \\\hline
{\small{\bb\mm}} & 53.7  & 1 & $0.3 \pm 0.06 \pm 0.1 $ & $74.6 \pm 0.7 \pm 0.7$\\\hline\hline
\multicolumn{5}{|c|}{\bf\boldmath $\sqrts=189$~GeV} \\\hline\hline
{\small{\bb\qq}}  & 172.1  &24  &   $19.9 \pm 0.8 \pm 2.9$ &  $47.0 \pm 0.8 \pm 1.6$ \\ \hline
{\small{\bb\nn}}  & 171.4 & 10 & $6.9 \pm 0.5 \pm 0.6$ &  $42.6 \pm 1.1 \pm 1.1$ \\\hline
{\small{\bb\tautau}} &  &  &  & $ 44.6 \pm 1.8 \pm
 2.0 $ 
 \\ \cline{1-1} \cline{5-5} 
{\small{\tautau\qq}} & \raisebox{1.5ex}[0cm][0cm]{168.7}   & \raisebox{1.5ex}[0cm][0cm]{3}  & \raisebox{1.5ex}[0cm][0cm]{$ 4.0\pm 0.5 \pm 0.9$}      &  $32.8 \pm 1.5 \pm 2.1$ \\\hline 
{\small{\bb\ee}} & 172.1 & 3 & $2.6 \pm 0.2 \pm 0.5$ & $66.3 \pm 1.1 \pm 1.3 $ \\\hline
{\small{\bb\mm}} & 169.4  & 1 & $2.1 \pm 0.1\pm 0.4$& $78.3 \pm 1.1 \pm 1.1$ \\\hline
\end{tabular}
\caption{\label{tab:smflow}\sl
  The \ho\Zo\  channels:
  the number of events for 
  the data, the total expected background normalised to the integrated 
luminosity of the data,
  and the detection efficiency for a Higgs boson decaying only into \bb\ or
\tautau, 
for typical masses close to the kinematical limits 
of 85 and 95 GeV at \sqrts~=~183 and 189 GeV, respectively. 
Two separate efficiencies are shown in the tau channel for the two
processes \ho\ra\bb, \Zo\ra\tautau\ and \ho\ra\tautau, \Zo\ra\qq.
The first error is statistical and the second systematic.}
\end{center}
\end{table}
\vspace*{-0.4cm}
\hspace{-0.75cm}
Higgs boson decays. Detailed descriptions of the flavour independent 
analyses are given 
in Sections ~\ref{sect:zhsearches} and ~\ref{sect:ahsearches}
for the processes \ee\ra\h\Z\ and \ee\ra\h\A, respectively.

The channels studied in ~\cite{pn183} and ~\cite{pr285}
using b-tagging, 
together with those looking for $\tau$--leptons, 
provide useful information in regions of 
the 2HDM(II) parameter space where the Higgs bosons are expected 
to decay predominantly into \bb\ and \tautau\ pairs. At $\sqrt{s} =$ 183 and 189 GeV, 
for the process \ee\ra\ho\Zo\, 
the following final states are considered:
\ho\Zo\ra\bb\qq, \bb\nn, \bb\tautau, \tautau\qq, \bb\ee\ and \bb\mm.
The 2HDM(II) process \ho\Zo\ra\Ao\Ao\Zo, followed 
by \Ao\ra\bb, is included when kinematically allowed. 
In addition, the 2HDM(II) associated production process, \ee\ra\Ao\ho, 
followed by \Ao\ho\ra\bb\bb, \Ao\ho\ra\bb\tautau\ (or \tautau\bb) and
\ho\Ao\ra\Ao\Ao\Ao\ra\bb\bb\bb, is studied.  

At \sqrts\ $\approx$ \mZ\ the following final states are 
interpreted in the 
framework of the 2HDM(II):
\ho\Zo\ra\qq\nn, \qq\tautau, \tautau\qq, 
\qq\ee\ and \qq\mm, 
as well as \Ao\ho\ra\qq\tautau\ (or \tautau\qq), 
and \ho\Ao\ra\Ao\Ao\Ao\ra\bb\bb\bb,   
if \ho\ra\Ao\Ao\ is kinematically allowed. 

The luminosity, the number of candidate events, the expected 
SM backgrounds,  and the efficiencies for each
of these \ho\Z\ and \ho\Ao\ channels at 183 and 189 GeV 
centre-of-mass energy are
given in Tables~\ref{tab:smflow} and \ref{tab:ahflow}, respectively.
The signal detection efficiencies for the process \ho\Zo\ra\Ao\Ao\Zo\   
can be found in ~\cite{pn183,pr285}.
\begin{table}[ht]
\begin{center}
\begin{tabular}{|c|c|c|c|c|}\hline
Channel \Ao\ho\ra &Luminosity [pb$^{-1}$] & Data & Total bkg. & Efficiency [\%]\\\hline\hline
\multicolumn{5}{|c|}{\bf\boldmath $\sqrts = 183$~ GeV} \\\hline\hline
{\small{\bb\bb}}  & 54.1   & 4    &  $2.92 \pm 0.2 \pm 0.5$  & $50.3 \pm 0.7 \pm 2.0$\\\hline
{\small{\bb\tautau}} & 53.7   & 3    &  $1.50 \pm 0.1 \pm 0.2$ & $44.7 \pm 1.6 \pm 1.8$\\\hline
{\small{\bb\bb\bb}}  & 54.1   & 2    &  $2.3 \pm 0.2 \pm 0.03$  & $36.0 \pm 2.16 \pm 1.8$\\\hline\hline
\multicolumn{5}{|c|}{\bf\boldmath $\sqrts = 189$~ GeV} \\\hline\hline
{\small{\bb\bb}}  & 172.1  & 8    &  $8.0 \pm 0.5 \pm 1.4$  & $48.4 \pm 0.7 \pm 3.9 $\\\hline
{\small{\bb\tautau}} & 168.7 & 7    &  $4.9 \pm 0.6 \pm 1.6$ & $45.3 \pm 1.5 \pm 2.3$\\\hline
{\small{\bb\bb\bb}}  & 172.1  & 5    &  $8.7 \pm 1.0 \pm 2.5$  & $45.4 \pm 2.2 \pm 4.3$\\\hline\hline
\end{tabular}
\end{center}
\caption{\label{tab:ahflow}\sl
         The \ho\Ao\  channels:
         the number of events for the data, the total expected background
         normalised to the integrated luminosity of the data and
         the signal efficiency for
         (\mh,~\mA)$=$(70~GeV,~70~GeV) and (80~GeV,~80~GeV) 
         in the \ho\Ao\ra\bb\bb\ and \bb\tautau\ 
         channels and 
         for (\mh,~\mA)$=$(60~GeV, 30~GeV) and (70~GeV, 20~GeV) in 
          the \ho\Ao\ra\Ao\Ao\Ao\ra\bb\bb\bb\  channel at 
         \sqrts = 183 and 189 GeV, respectively.
         The first error is statistical and the second systematic.  
}
\vspace*{0.4cm}
\end{table}
The detection
efficiencies quoted in Tables~\ref{tab:smflow} and ~\ref{tab:ahflow}
are given as an example for specific values of \mh\ and \mA.
When scanning the parameter space the efficiency 
is calculated for each point in the (\mh,~\mA)
plane for each of the final states considered.
In the tau channel two different final state topologies are studied,
in which \ho\ra\bb, \Zo\ra\tautau\ and \ho\ra\tautau, \Zo\ra\qq,
providing different efficiencies, as shown in Table~\ref{tab:smflow}.
The dominant contributions to the systematic errors on the 
signal and background efficiencies come from the 
uncertainty related to b--tagging ~\cite{pn183,pr285}.

\section{Flavour independent searches for ${\protect \boldmath \ee\ra\h\Z}$}
\label{sect:zhsearches}
\vspace*{0.3cm}

This section describes the searches for 
the \ee\ra\h\Z\ process at \sqrts = 189 GeV in the following final states: \h\Z\ra\qq\qq\ and
\glgl\qq\ (the four--jet channel), \h\Z\ra\qq\nn\ and \glgl\nn\ (the
missing--energy channel), \h\Z\ra\qq\tautau\ and \glgl\tautau\ (the
tau channel), \h\Z\ra\qq\ee\ and \glgl\ee\ as well as \h\Z\ra\qq\mm\
and \glgl\mm\ (the electron and muon channels).  
A new flavour independent selection has been developed 
for the four--jet channel.
The other channels follow closely the analyses described in reference 
\cite{pr285} but do not make use of b--tagging information.

\vspace*{0.4cm}
\subsection{\label{sect:4jet} The four--jet channel}

The search in the four--jet channel is a {\it{test mass dependent analysis}}
using a binned maximum likelihood method. In order
to obtain high sensitivity over a wide range of
Higgs boson masses, several likelihood analyses are performed. Each of
these is dedicated to test a specific Higgs mass hypothesis. The
test masses are chosen from \mh $=60$~\G\ up to \mh $=100$~\G\ in steps
of $1$~\G , significantly less than the expected mass
resolution for the Higgs boson, which is $2$ to $3$~\G. Signal Monte
Carlo events have been generated and reconstructed for each
of these test masses. Each likelihood is defined by reference
histograms made from the background Monte Carlo samples and the signal
Monte Carlo events generated at the corresponding mass. All data
events are then subjected to each of the $41$ resulting likelihood
analyses, and those passing the selection are counted as candidates
for masses in a window of $\pm$0.5~\G\ centered on the
respective test mass. A single event can be a candidate at 
a variety of different test mass
values, and the candidates found in the data are
not identical for all mass hypotheses between $60$ and $100$~\G.
\newpage
Correct assignment of particles to jets plays an essential role in
separating one of the main backgrounds, \WW\ra\qq\qq, from the signal
process, as well as in accurately reconstructing the mass of Higgs bosons
in signal events. The jet reconstruction method is explained in \cite{pr285}.
The initial preselection, designed to retain only events with four distinct
jets, is unchanged with respect to the four--jet channel 
in \cite{pr285} and is independent of any mass
hypothesis.

The following selection criteria and the
likelihood make explicit use of the mass hypothesis:
\begin{itemize}
\item[(1)]{A kinematic fit which is applied to test the \eerhz\ hypothesis 
imposes energy and momentum conservation. Additionally one of the dijet
masses is constrained to \mZ\ within its natural width and the other dijet 
mass is constrained to the tested mass of the Higgs boson (ZH--Fit). 
This fit is applied to all six possible jet
associations and for at least one of these combinations it is
required to converge with a $\chi^2$--probability larger than
10$^{-5}$. The candidate mass is later calculated using the jet 
association which yields the highest $\chi^2$--probability.}
\item[(2)]{To discriminate against \WW\ background, the ratio of the
\me\ probability for Higgs--strahlung \cite{higgsme} and the \me\
probability for \WW\ production as implemented in
EXCALIBUR~\cite{excalibur} is required to be larger than $1.2\times 10^{-4}$.
When calculating the \me\ probability for Higgs--strahlung it is
necessary to assign the measured jets to the original partons. Which
jets belong to the decay products of the \Z\ and which to the Higgs is
determined by the best $\chi^2$--probability of the kinematic fit
described in criterion (1). As it remains unknown whether the \Z\ decays into
up--type or down--type quarks and which jet belongs to the quark and
which to the anti--quark, the \me\ is averaged over all these
combinations.  Also, the \me\ probability for \WW\ production is
averaged over all possible jet-parton assignments.
For both matrix element probabilities the four--vectors after a four--constraint (4C) fit, imposing energy and momentum conservation, are used as input.}
\end{itemize}

The following six variables are combined with
a binned likelihood method~\cite{smpaper172}, with one class for the
signal and two for the 4-fermion and the 2-fermion backgrounds:
\begin{itemize}
\item[(a)]{The logarithm of the ratio of the \me s for Higgs--strahlung 
(ME$_{\mathrm{ZH}}$) and for \WW\ production (ME$_{\mathrm{WW}}$).}
\item[(b)]{The logarithm of the \me\ probability for the Higgs--strahlung 
process. In contrast to the
kinematic fits, the \me\ also contains angular information, which 
allows one to distinguish kinematically between \ee\ra\ho\Zo\ 
and \ZZ\ production, if \mh\ is in the region of \mZ.
While variable (a) mainly discriminates against \WW\ events,
variable (b) helps to select events compatible with a signal hypothesis.
The correlations between (a) and (b) are small.}
\item[(c)]{The logarithm of the $\chi^2$--probability of
the ZH--Fit to the Higgs--strahlung hypothesis of selection criterion (1).
Only the jet association that gives the highest fit probability is
considered.}
\item[(d)]{The logarithm of the $\chi^2$--probability resulting from a
kinematic fit (WW--Fit), which in \thinspace addition to energy and \thinspace momentum conservation
forces both \thinspace dijet masses to \thinspace be 
\begin{center}
\begin{figure}[H]
\epsfig{file=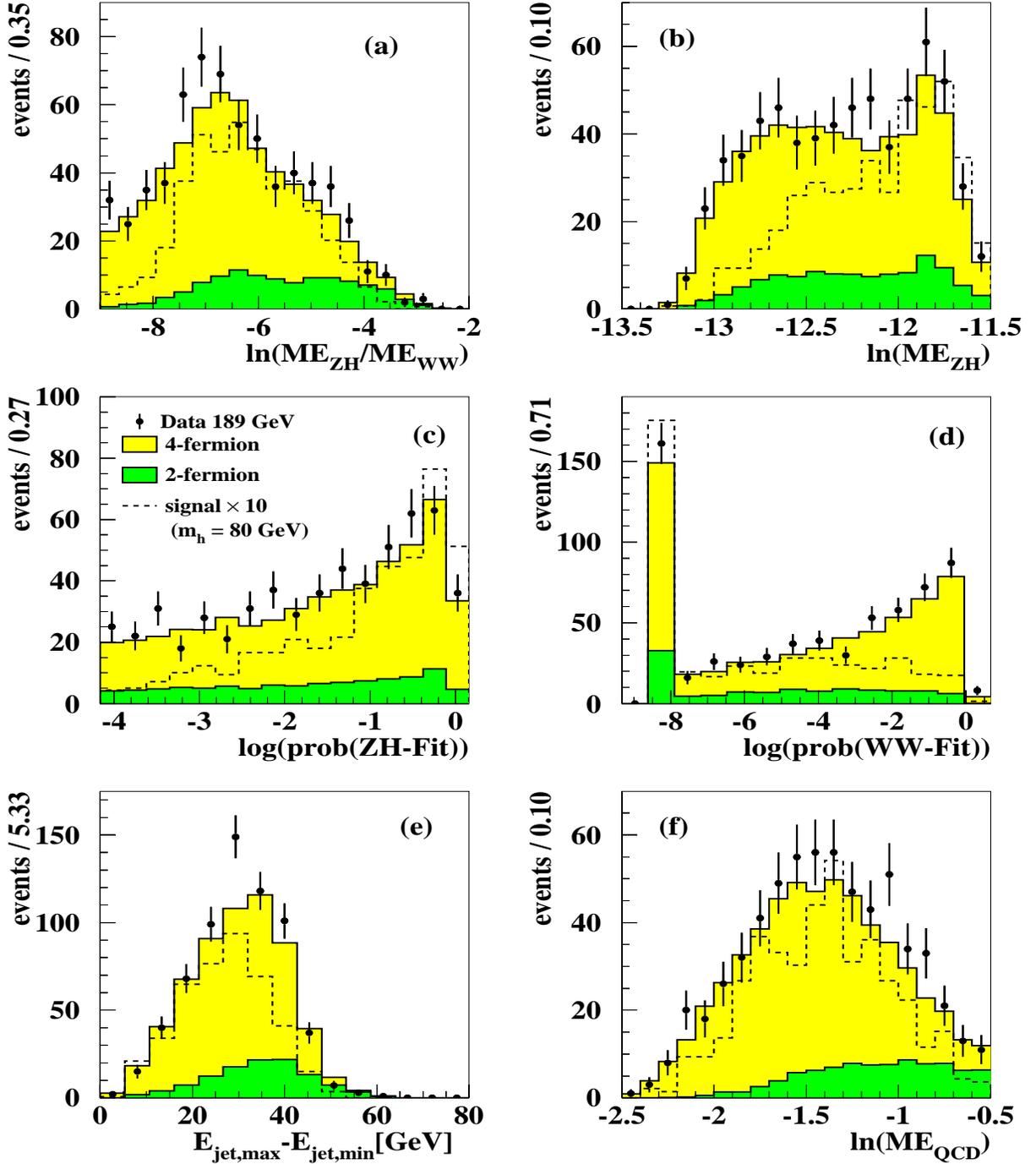,width=16.8cm,height=20.0cm}
\caption[]{\label{fig:zhqqlike1}\sl
Input variables of the four--jet channel likelihood selection for
\mh$=80$~\G. The OPAL data are indicated by dots with error bars
(statistical errors), the 4-fermion background by lighter grey
histograms, and the 2-fermion background by darker grey
histograms. All MC distributions are normalised to the integrated
luminosity of the data. The estimated contribution from an 80~\G\ SM
Higgs boson, scaled up by a factor of 10, is shown with a dashed line in each case.}
\end{figure}
\end{center}
\newpage
\begin{center}
\begin{figure}[H]
\epsfig{file=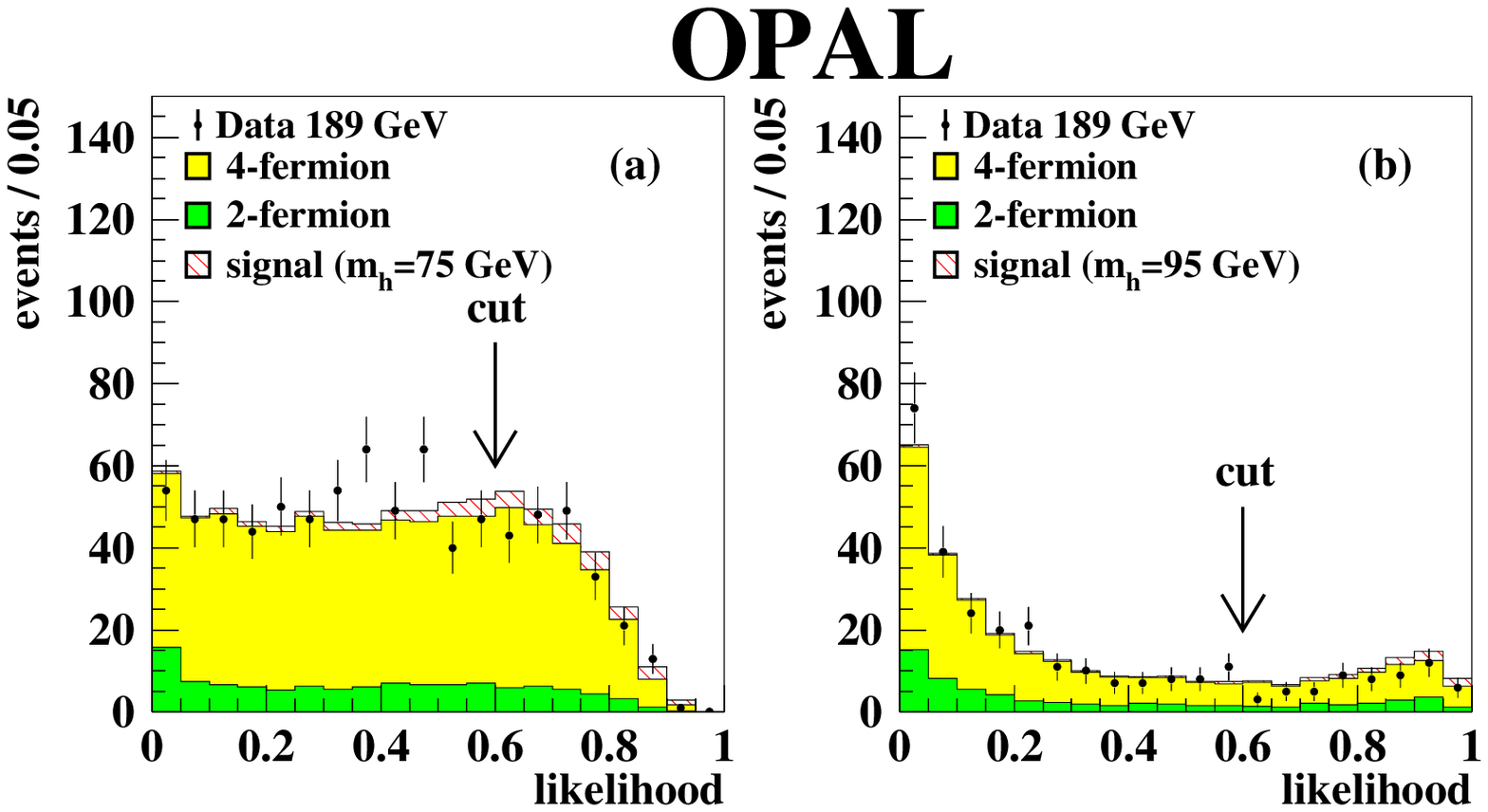,width=17.0cm,height=10.0cm,width=18.0cm}
\caption[]{\label{fig:zhqqlike}\sl
Likelihood distribution of the four--jet channel for test masses of
(a) $75$~\G\ and (b) $95$~\G. The OPAL data are
indicated by dots with error bars (statistical errors), the 4-fermion
background by lighter grey histograms, and the 2-fermion background
by darker grey histograms. The contributions expected from a (a) 75~\G\
and (b) 95~\G\ \ho\ boson assuming SM cross--section and
branching ratios are shown as hatched histograms. All MC distributions
are normalised to the integrated luminosity of the data. All events
with a likelihood larger than 0.6 are accepted.}
\end{figure}
\end{center}
equal to the mass of the W boson. Only the jet association that gives
the highest fit probability is considered.}
\item[(e)]{The difference between the largest and the smallest jet
energies in the event.}
\item[(f)]{The logarithm of an event weight, ME$_{\mathrm{QCD}}$, formed
\cite{qcdop} from the tree level matrix element for the process
\ee\ra\qq\qq, \qq\glgl\ \cite{qcdme}, to reduce QCD background.}
\end{itemize}
In Figure~\ref{fig:zhqqlike1} the distributions of the likelihood 
input variables are shown for the difficult 
case in which \mh $\approx$ \mWn. 

\begin{figure}[H]
\begin{center}
\epsfig{file=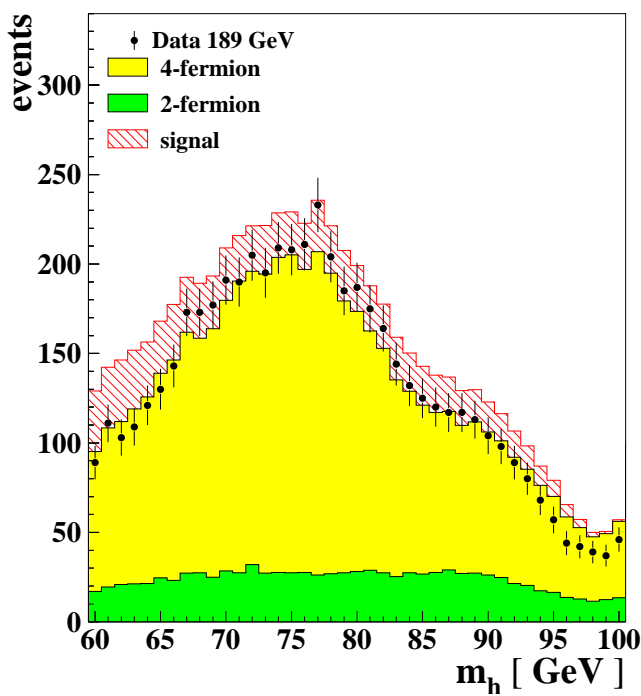,width=10.0cm}
\caption[]{\label{fig:zhqqcand}\sl
Number of events that pass the four--jet channel selection at each
test mass. The number of candidates found in the data is indicated by
dots with error bars (statistical errors), and the darker and lighter grey 
histograms
correspond to the number of events expected from 2-fermion and
4-fermion backgrounds, respectively. The expected contribution from a
Higgs boson with a mass equal to the test mass, assuming SM cross--sections 
and branching ratios, is shown by the hatched histogram. 
All MC distributions are
normalised to the integrated luminosity of the data.
Each bin corresponds to a different analysis of the same data,
leading to a strong correlation between neighbouring bins.
}
\end{center}
\end{figure}

The distributions of the final likelihood
${\cal L}^{\mathrm hZ}$ are shown in Figure~\ref{fig:zhqqlike}
for test masses of $75$ and $95$~\G. Candidates for signal production
are required to have ${\cal L}^{\mathrm hZ}>0.6$ for all test masses. 
In Table~\ref{tab:smflowj} the numbers of observed and
expected events, together with the detection efficiencies for a Higgs
signal with a mass of 80~\G , are given.
Figure~\ref{fig:zhqqcand} compares the number of candidate
events obtained after the likelihood selections for the different
test masses with the expected background evaluated from Monte Carlo
simulations. In the region of 75~\G, the number of expected background
events rises to more than 200. This is due to the presence of 
\WW\ events, where the
mass of one of the W bosons is constrained to the mass of the \Z,
which reduces the other dijet mass by a few

\begin{center}
\begin{figure}[H]
\epsfig{file=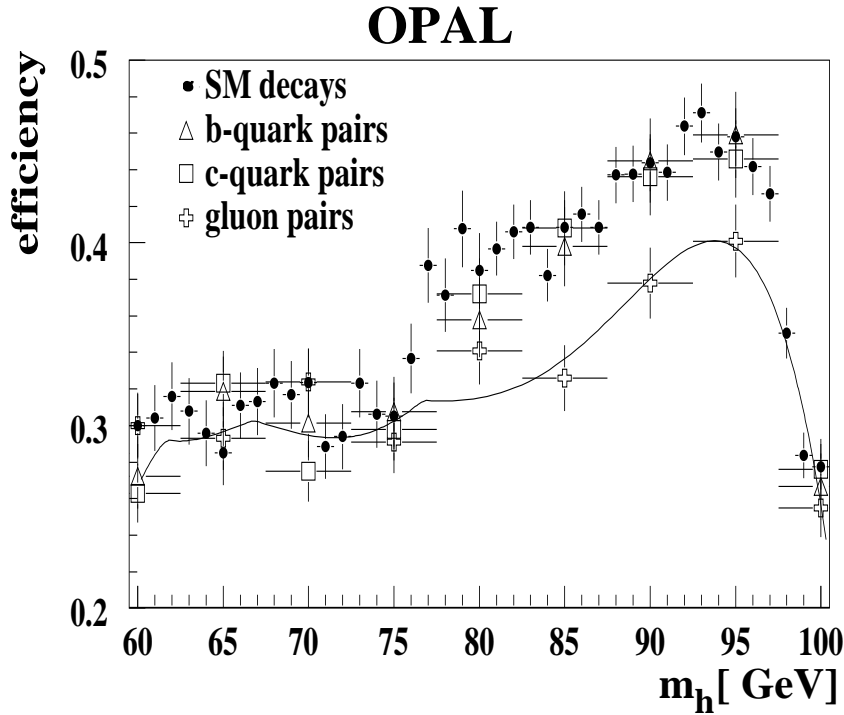,width=13.0cm}
\caption[]{\label{fig:zhqqeff}\sl
Efficiency of the four--jet channel selection as a function of \mh\,
determined from different Monte Carlo samples. The dots represent the
efficiency obtained from samples with SM branching ratios
which also define the reference histograms at each test mass. The
triangles, squares and crosses correspond to independent samples where
the Higgs boson decays exclusively to b--quark, c--quark or gluon pairs,
respectively. For the limit calculation, efficiencies have been fitted
to a polynomial function of \mh\ for each flavour separately, and at each
mass point the lowest fitted polynomial is used (black line).} 
\end{figure}
\end{center}

\hspace{-0.75cm}
\G\ compared to the
nominal value. The candidate masses are calculated from the momenta 
resulting 
from a 5C fit requiring energy and momentum conservation and forcing
one of the dijet masses to \mZ. 
Figure~\ref{fig:zhqqeff} shows the efficiency as a function of the Higgs
mass for decays to b--quark, c-quark and gluon pairs separately as well
as for a mixed sample according to SM branching ratios. For \mh\
between $80$ and $95$~\G\ the efficiency reaches about $40$ to $45\%$
for quarks and $35$\% for gluons. For small values of \mh\ it drops to
$25\%$ due to the relatively large amount
of initial state radiation that accompanies Higgs--strahlung 
when \mh\ is considerably lower than the kinematic limit, \ie\, \sqrts$-$\mZ.
For the limit calculation, the
efficiencies have been fitted to a polynomial function 
of \mh\ for each flavour separately, and at each
mass point the lowest fitted polynomial is used. 
The signal selection efficiencies are affected by the 
uncertainties given below, expressed in relative percentages and 
shown as an example for a Higgs mass of $75$~\G. 
The systematic errors have been evaluated as follows: for each variable,
the distributions obtained from the background Monte Carlo samples
have been compared to those in the data. Then all Monte Carlo events
(including the signal samples) have been reweighted so that the mean
values of data and background distributions become the same and so that
the sum of all

\hspace{-0.75cm}
weights is equal to $1$. The relative deviations in the
number of events which pass the selection obtained by reweighting
according to the different variables have been added in quadrature and
amount to 5.5\%. 
The same procedure has been applied to the kinematic
likelihood variables, yielding an uncertainty of
2.0\%. The 
uncertainty on the error parameterisation of the jet
momenta used in the kinematic fits has been evaluated by varying the
energy and angular resolutions by $\pm$10\% , the energy scale by
$\pm$ 1\% and the centre--of--mass energy by $\pm$0.3~\G , each time
repeating all kinematic fits. This leads to an uncertainty of
6.4\%. Since the steps in
the test mass are chosen to be smaller than  
the expected mass
resolution, the deviation in efficiency due to the interpolation
between test masses amounts only to 0.7\%. The total systematic
uncertainty on the signal selection efficiency has been calculated by
adding the above sources in quadrature yielding 8.7\%. All of these error
contributions have been evaluated for masses of $60$ and
$90$~GeV as well, leading to total systematic uncertainties of 9.8\% and 7.1\%,
respectively. The Monte Carlo statistical error is about 2\%.

The following uncertainties on the two major background sources are
taken into account (the first number corresponds to the
4-fermion, the second number to the \qq($\gamma$) background, both for a
test mass of $75$~\G): the uncertainty from modelling of the
preselection cuts is evaluated as described above and amounts to
5.2\%/4.5\%. For the \thinspace likelihood \thinspace variables \thinspace this procedure \thinspace leads \thinspace to an \thinspace
uncertainty \thinspace of 1.2\%/1.6\%. \thinspace Varying the 
\begin{center}
\begin{figure}[H]
\epsfig{file=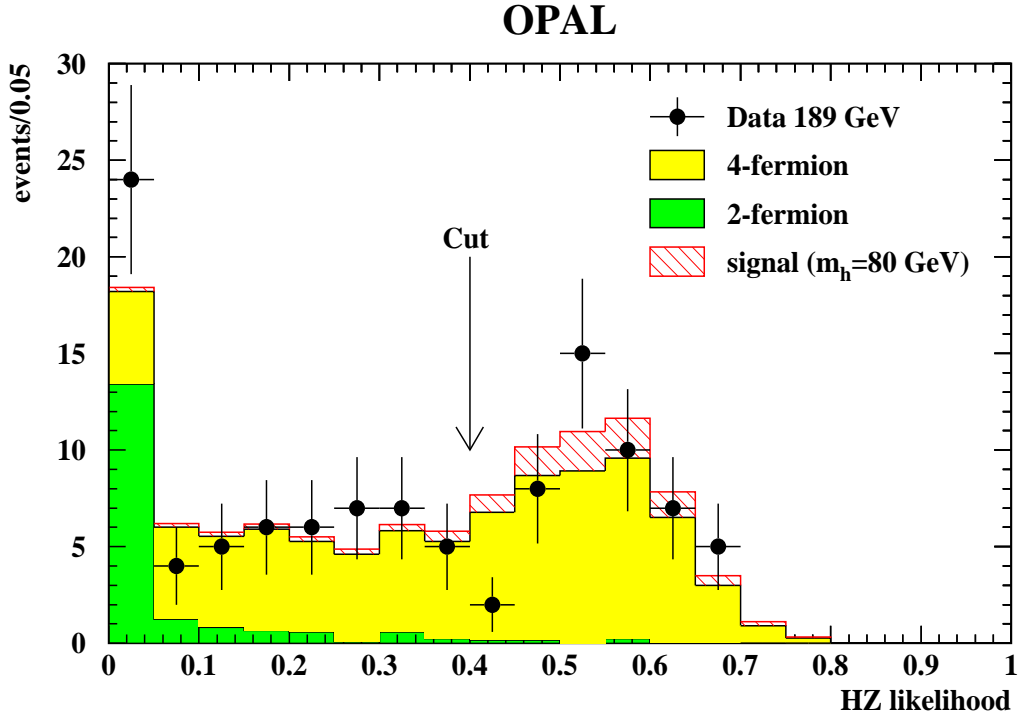,width=15.0cm,height=10.8cm}
\caption[]{\label{fig:misslike}\sl
Likelihood output of the missing energy channel selection. The OPAL
data are indicated by dots with error bars (statistical errors), the
4-fermion background by the lighter grey histogram, and the
2-fermion background by the darker grey histogram. The estimated
contribution from an 80~\G\ Higgs with SM cross--section and branching ratios
is shown as a
hatched histogram. All MC distributions are normalised to the
integrated luminosity of the data.}
\end{figure}
\end{center}

\hspace{-0.75cm}
error parameterisations,
energy scale and centre--of--mass energy for the kinematic fits yields
an uncertainty of 3.0\%/8.1\%. Different Monte Carlo generators have
been used to evaluate the background from 4-fermion events
(EXCALIBUR instead of grc4f) and QCD events (HERWIG instead of PYTHIA)
resulting in an uncertainty of 2.2\%/12.3\%. The total systematic
uncertainty on the residual background estimate amounts to 10.3\%. For
test masses of $60$ and $90$~\G\ the total systematic uncertainties
amount to 12.4\% and 8.3\%, respectively. The largest Monte Carlo statistical
error is 1.3\%. 
\subsection{The missing--energy channel}
\label{sect:sm-miss}
This analysis is nearly identical to a previous one, of which a detailed description
can be found in~\cite{pn183}, with the exception
that the b--tagging is not applied. The preselection
is unchanged.  The same
kinematic variables used previously, as well as the acollinearity angle
and the total missing transverse momentum, are
combined using a likelihood technique. In Figure~\ref{fig:misslike}, the resulting
signal likelihood distribution is shown for the data, SM backgrounds,
and an example signal at $m_{\mathrm h}=80$~GeV. The signal
likelihood is required to be larger than 0.4 for an event to be
selected as a Higgs candidate.

\vspace*{0.5cm}
\begin{center}
\begin{figure}[H]
\epsfig{file=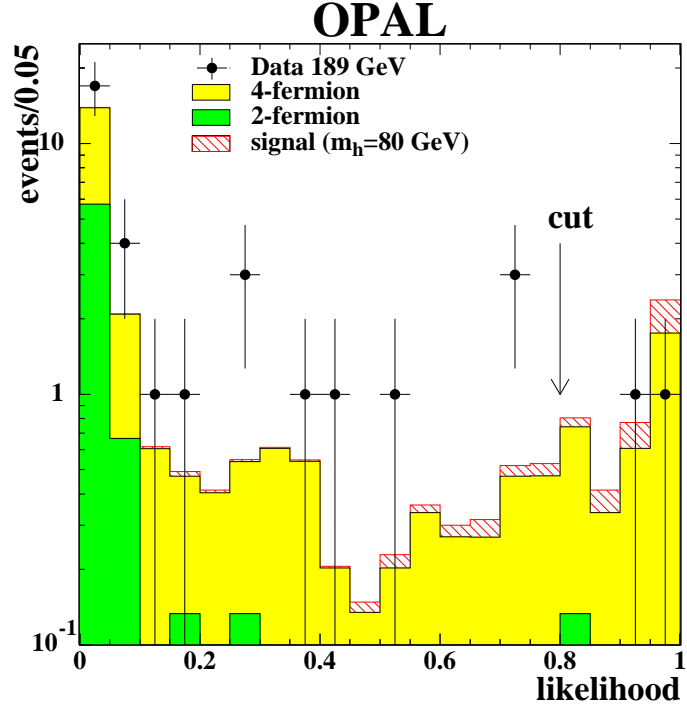,width=10.0cm,height=10.0cm}
\caption[]{\label{fig:tauinputs}\sl
Likelihood output of the tau channel selection for events
satisfying the \ho\Zo\ra \qq\tautau\ kinematic hypothesis.  OPAL data are
indicated by dots with error bars (statistical errors), 
4-fermion background by the lighter
grey histogram, and 2-fermion background by the darker grey histogram.
The contribution expected from an 80 GeV Higgs with SM cross--section and
branching ratios is shown as a hatched histogram.  All MC distributions
are normalised to the integrated luminosity of the data.}
\end{figure}
\end{center}

\vspace*{0.8cm}
The reconstructed mass in selected events is evaluated using a
kinematic fit constraining the recoil mass to the \Z\ mass. The
numbers of observed and expected events\footnote{In the calculation
of the efficiencies and backgrounds in the missing--energy channel, a
2.5\% relative reduction has been applied to the Monte Carlo estimates
in order to account for accidental vetoes due to accelerator related
backgrounds in the forward detectors.} are 
given in
Table~\ref{tab:smflowj}, together with the selection efficiencies for
an 80 GeV Higgs.  The selection efficiency has been estimated from
Monte Carlo samples generated separately for b--quark, c--quark and
gluon pairs. The efficiency is the lowest and
the width of the reconstructed Higgs mass is the 
largest for c--quark pairs and 
therefore they are used in the limit calculation. For an 80 GeV Higgs,
the efficiency is (51.0$\pm$1.7(stat.)$\pm$1.3(syst.))\%.  
The efficiency has 
been optimised for Higgs masses between 70 and 90~\G. Outside this
range, the efficiency decreases and reaches about 20\% for masses of
60 and 100~\G. For b--quarks, the efficiency is about 5\% 
(relative) higher throughout the whole mass range.
A total of 47 data events pass the selection, while
44.5$\pm$1.4(stat.)$\pm$3.0(syst.) events are expected from SM
background processes.
The systematic error has been evaluated as in~\cite{pn183},
but b--tagging related errors have been omitted.

\subsection{The tau channel}
\label{tau}
\vspace*{0.3cm}
The preselection, the tau lepton identification
using an artificial neural network, and the two--tau likelihood,
$\mathcal{L}_{\tau\tau}$, used in this channel are unchanged
with respect to the analysis described in reference ~\cite{pr285}.
Since b--tagging information is not used, for the final selection the
likelihood $\mathcal{L}( \qq\tautau)$ ~\cite{pr285} is used
and required to exceed 0.8. Additionally, the $\chi^2$--probability
of a kinematic fit, which constrains the invariant mass of the two
$\tau$'s to \mZ, should be larger than $10^{-5}$, since
this analysis is designed to be sensitive to 
hadronic Higgs boson decays and to \Zo\ra\tautau.
In Figure~\ref{fig:tauinputs} the resulting
likelihood distributions are shown for the data, SM backgrounds,
and an example signal at $m_{\mathrm h}=80$~GeV. 
The numbers of observed and expected events 
are given in Table~\ref{tab:smflowj}, together with the
selection efficiencies for an 80 GeV Higgs. The signal detection
efficiency has been evaluated for b--quark, c--quark and gluon pairs
separately and at each mass the lowest value is taken for the limit
calculation. For an 80~\G\ Higgs boson it amounts to (28.7 $\pm$
1.5(stat.)$\pm$ 2.7(syst.))\% after all the selection
requirements. For lower and higher Higgs masses, the efficiency drops
to about 20\% at \mh$=50$~\G\ and \mh$=100$~\G. Two events survive the
likelihood cut, which can be compared to the expected background of
$3.4 \pm 0.5 (\mrm{stat}.) \pm 0.7 (\mrm{syst}.)$.  
The systematic errors quoted above for signal and 
background are evaluated with the
method described in~\cite{pn183}, with the contributions from
fragmentation and decay multiplicity of b--quarks omitted. 

\vspace*{0.6cm}
\subsection{The electron and muon channels}
\vspace*{0.3cm}

\begin{center}
\begin{figure}[H]
\epsfig{file=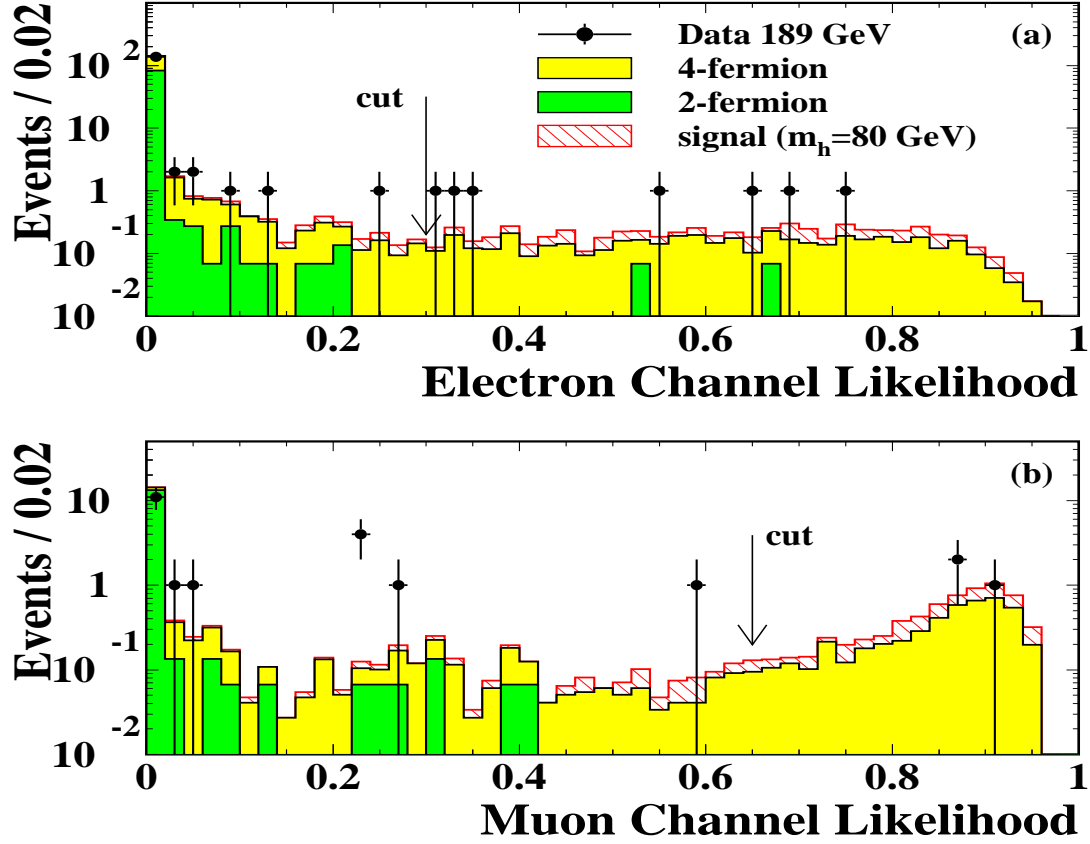,width=16.0cm,height=14.cm}
\caption[]{\label{fig:leptonic}\sl
(a) The electron and (b) the muon channel likelihood distributions. The OPAL
data are indicated by dots with error bars (statistical), the 4-fermion
background by lighter grey histograms, and the 2-fermion background
by darker grey histograms.
The contributions expected from an 80~\G\ Higgs with SM cross--section 
and branching ratios are
shown as hatched histograms. All MC distributions are
normalised to the integrated luminosity of the data. }
\end{figure}
\end{center}

The preselection cuts and kinematic likelihood $\cal K$ are
identical to the analysis described in ~\cite{pn183}. 
Because the present analysis is intended to be independent of the
flavour of the Higgs decay products, no b--tagging is applied and $\cal
K$ is used as the final selection variable, which should exceed 0.3
for the electron and 0.65 for the muon
channel. Figure~\ref{fig:leptonic} shows the distribution of 
$\cal K$ for the electron
(a) and the muon (b) channel. 

The signal selection efficiency has been evaluated and fitted for each
flavour separately. For the limit calculation, the
lowest of these efficiencies at each mass point has been used.
For an 80~GeV Higgs boson
it amounts to (55.4$\pm$1.6(stat.)$\pm$0.6(syst.))\% for the electron
channel, and (59.3$\pm$1.5(stat.)$\pm$0.7(syst.))\% for the muon
channel. In the electron channel, the efficiency lies between 50\% and
55\% for all Higgs masses between 60 and 95~\G, and drops to 27\% for
a Higgs mass of 100~\G. In the muon channel, the efficiency is between
55\% and 70\% for all Higgs masses under consideration.

The numbers of observed and expected events  
are given in Table~\ref{tab:smflowj}, together with the
detection efficiency for an 80~GeV SM Higgs boson. The selection
retains 7 events in the electron channel and 3 in the muon
channel. The total background expectation is
4.7$\pm$0.2(stat.)$\pm$1.4(syst.) events in the electron channel and
4.7$\pm$0.1(stat.)$\pm$0.9(syst.) events in the muon channel.

The systematic errors quoted above for signal and background are
evaluated with the method described in~\cite{pn183}. The largest
contribution to the systematic error on the background expectation is
due to differences between various Monte Carlo generators. 
\subsection{Summary of the flavour independent searches for \\ 
${\protect \boldmath \ee\ra\h\Z}$}
\label{subsection:modindep}

The total numbers of candidates accepted after the preselection and the
flavour independent likelihood cut, 
compared to the expected SM backgrounds as well as the detection 
efficiencies for a hadronically decaying Higgs boson with a mass of 80~\G, 
are summarised in Table~\ref{tab:smflowj}, together with the expected number
of signal events in the 2HDM(II) for the case of
$\alpha$~=~0, \tanb~=~1.0 and \mh~=~80 GeV. 
The total number of observed candidates for an 80~\G\ Higgs boson is 246, 
while the background expectation amounts to 
$231.3\pm 4.4({\mathrm stat.}) \pm 18.0({\mathrm syst.})$.
{\small{
\begin{table}[t]
\vspace*{-1.0cm}
\begin{center}
\begin{tabular}{|c||r||r||r|r||c||c|}\hline
Cut & Data & Total bkg. & q\=q($\gamma$) & 4--fermion  &
    Efficiency [\%] & Signal \\ \hline\hline
\multicolumn{7}{|c|}{Four--jet Channel ~~ ${\cal L}$ = 174.1 pb$^{-1}$} \\\hline
Preselection & 1568 & 1521 & 379 & 1142 & 87.4 & 31.8 \\\hline (1) & 696 & 649 & 117 &
532 & 65.3 & 23.8\\ (2) & 648 & 600 & 116 & 484 & 64.1 & 23.3 \\\hline
Likelihood & 187 & 174 & 28 & 146 & 35.4 & 12.9 \\\hline
\hline
\multicolumn{7}{|c|}{Missing--energy Channel ~~ ${\cal L}$ = 171.8 pb$^{-1}$} \\\hline 
Preselection & 111 & 101 & 18 & 83 & 63.4 & 6.5\\\hline 
Likelihood & 47 & 44.5 & 0.6 & 43.9 & 51.0 & 5.3\\\hline\hline
\multicolumn{7}{|c|}{Tau Channel ~~ ${\cal L}$ = 168.7 pb$^{-1}$}  \\ 
\hline
Preselection & 185 & 156 & 55 & 101 & 49.1 & 0.8\\\hline
Likelihood & 2 & 3.4 & 0.1 & 3.3 & 28.7 & 0.5 \\\hline\hline
\multicolumn{7}{|c|}{Electron Channel ${\cal L}$ =  172.1 pb$^{-1}$} \\\hline\hline
Preselection & 152 & 153 & 84 & 69 & 77.8 & 1.4 \\\hline
Likelihood & 7 & 4.7 & 0.1 & 4.5 & 55.4 & 1.0 \\\hline\hline
\multicolumn{7}{|c|}{Muon Channel ${\cal L}$ = 169.4 pb$^{-1}$} \\\hline\hline 
Preselection & 22 & 22 & 14 & 8 & 78.8 & 1.3 \\\hline 
Likelihood & 3 & 4.7 & 0.0 & 4.7 & 59.3 & 1.0\\\hline
\end{tabular}
\caption{\sl The \h\Z\ channels for the flavour independent analysis: 
the numbers of events selected in the
data at \sqrts~=~189 GeV, expected background from SM processes normalised to the data
luminosity (given for each channel),
the minimal detection efficiencies for a Higgs
boson with a mass of 80~\G\ decaying to quark or gluon pairs and the
expected number of signal events within the 2HDM(II) for the case of
$\alpha$~=~0, \tanb~=~1.0 and \mh~=~80 GeV,
after the preselection and the flavour independent likelihood cut.
In the case of the four--jet channel, after preselection, the
numbers of events are given for the selection with an 80 GeV test mass.  
}
\label{tab:smflowj}
\end{center}
\end{table}}}
\begin{center}
\begin{figure}[H]
\vspace*{-0.5cm}
\centerline{
\mbox{\epsfig{file=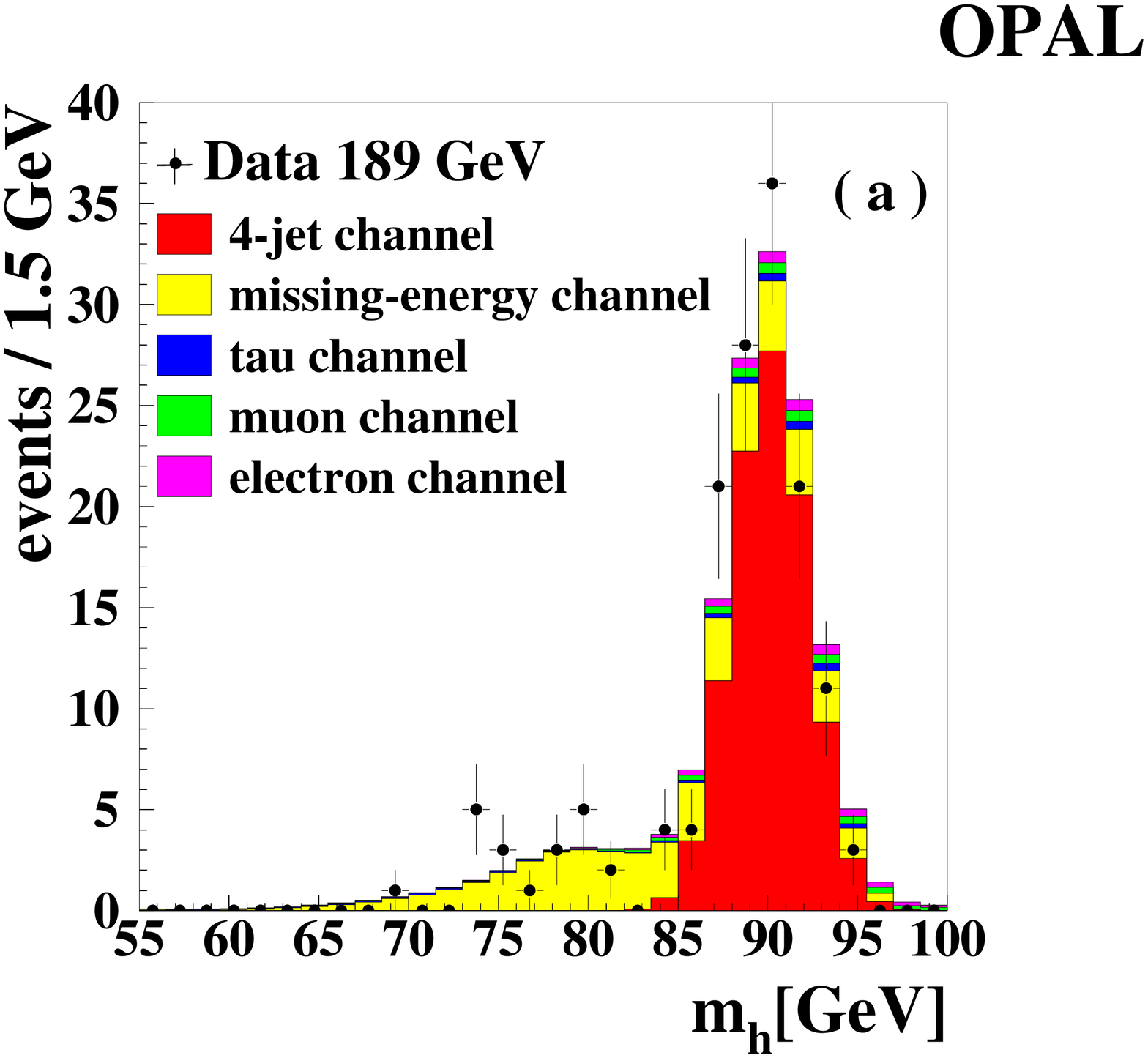,width=9.0cm}}\hfil
\mbox{\epsfig{file=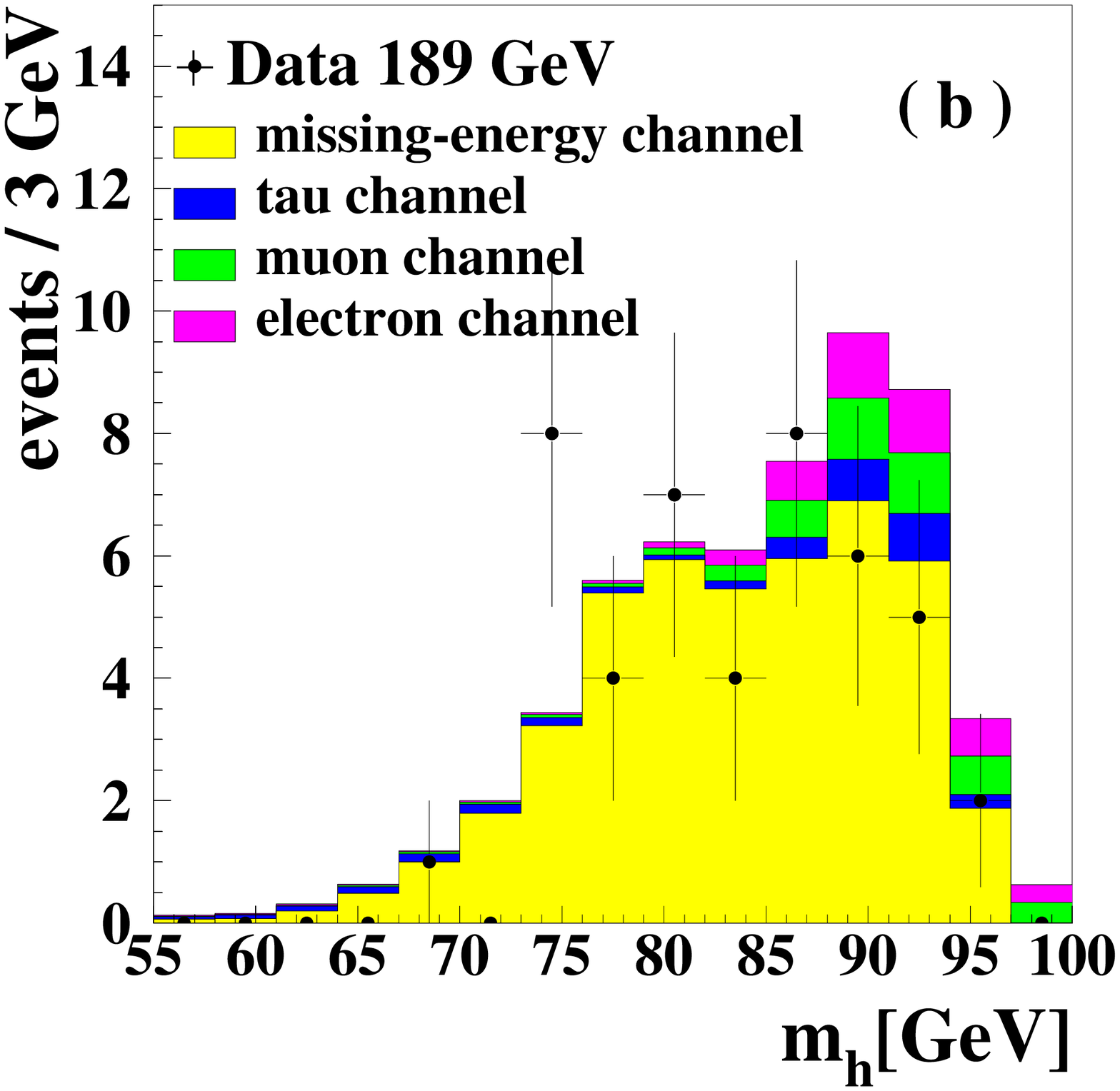,width=8.3cm}}}
\caption[]{\label{fig:massplot}\sl 
Mass distributions for selected data events and SM background
expectation (a) for all channels and 
(b) for all channels except the four--jet channel. 
OPAL data are indicated 
by dots with error bars (statistical) while the
background contributions from the various channels are shown in
different shades of grey. For the four--jet channel the 90~\G\ analysis
is shown.}
\end{figure}
\end{center}

\hspace{-0.75cm}
The mass distributions for candidates
found in the data as well as for the background expected from the SM
for all flavour independent channels are shown in Figure \ref{fig:massplot}(a).
The same mass distributions without the four--jet channel 
contribution are shown in Figure \ref{fig:massplot}(b).

\section{Flavour independent search for 
${\protect \boldmath\ee\ra\h\A}$}
\label{sect:ahsearches}

We have searched for the process
\ee\ra\h\A\ in the final states \qq\qq, 
\glgl\qq\ and \glgl\glgl.
The signal is characterised by events with four well-separated jets
with characteristic invariant masses of the dijet pairs
originating from the \h\ and the \A.
The dominant background is from the process
\ee\ra\Wp\Wm\ra\qq\qq.
The second-largest contribution 
comes from \ee\ra\Z$/\gamma$\ra\qq\ with multiple hard gluon
radiation producing a four--jet final state.

This search is designed to be sensitive over a large portion of the 
(\mh,~\mA) plane, and the kinematic signatures of signal events depend
strongly on \mh\ and \mA.  For this reason, a loose selection is
performed first, retaining four-jet hadronic events with partial rejection of
\Wp\Wm\ and Z$^0$Z$^0$ events.  The main
discrimination, however, is achieved by constraining candidate
events to the signal mass hypothesis (\mh\ and \mA),
and using the logarithm of the resulting $\chi^2$ as the discriminant
variable in the limit calculation instead of the reconstructed
dijet masses.  This choice of variable also incorporates naturally the
measurement uncertainties on the reconstructed masses and simplifies
the interpolation of its shape as a function of \mh\ and \mA.

\subsection{Selection}

\begin{figure}[t]
\begin{center}
\epsfig{file=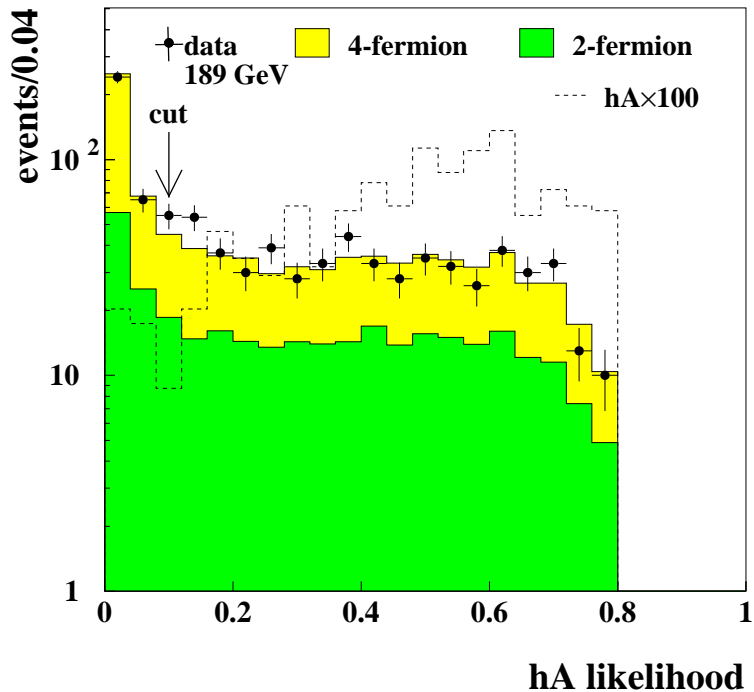,width=10.0cm}
\caption{\label{fig:taglesshalike}
\sl{Likelihood distribution for the  
flavour independent \ee\ra\h\A\ channel. The OPAL data are 
indicated by dots with error bars (statistical error),
the 4-fermion background by the lighter grey histogram, 
the 2-fermion background by the darker grey histogram 
and a representative signal for \mh\ = 30~GeV, \mA\ = 60~GeV by the dashed histogram.
All Monte Carlo distributions are normalised to the data luminosity and
the signal is scaled by a factor of 100. 
}}
\end{center}
\end{figure}

Candidate events must first satisfy the requirements of
a preselection and then a loose selection based on a likelihood variable
which is built out of reference distributions of reconstructed
quantities for events passing the preselection.
The following criteria are applied ((1)--(4) preselection,
(5) selection):
\begin{itemize}
\item[(1)]{Each candidate event is required to be classified as a hadronic
final state~\cite{hadronic} with an effective centre-of-mass
energy $\sqrt{s^\prime}$ exceeding 150~GeV.  
The jet resolution parameter in the Durham 
scheme~\cite{drm} $y_{34}$ is required to be larger than 0.003 in order to select
events with four distinct jets.}
\item[(2)] {Each jet must have at least three tracks.}
\item[(3)] {The $\chi^2$ probability of a 4C fit, requiring energy 
and momentum conservation, must be greater than $10^{-5}$,
to ensure that the mass reconstructions
used to isolate the signal do not suffer from poor measurement or
energy loss from initial state radiation.
}
\item[(4)]{A 6C kinematic
fit is performed requiring energy and momentum conservation and
also that the invariant masses of the dijet pairs are equal to
\mWn.
The 6C fit $\chi^2$ probability of each of the
three possible jet combinations is required to be less than 0.01,
to reduce the background from hadronic \Wp\Wm\ decays.}
\item[(5)]{ 
A likelihood composed of five variables is computed.
These variables are the jet resolution parameter $y_{34}$, 
the event-shape variable $C$ obtained from the
eigenvalues of the sphericity tensor~\cite{cparam},
the smallest angle between any two jets in the event, the logarithm of
the QCD matrix element ME$_{\mathrm{QCD}}$~\cite{qcdme},
and the largest $\chi^2$ probability
of three 5C kinematic fits constraining
energy and momentum and requiring the equality of the masses of the two
dijet systems (three possible pairings).
The QCD matrix element used is the maximum 
of the matrix elements considering
all possible assignments of observed jets to partons in the
\ee\ra\qq\qq\ and \ee\ra\qq\glgl\ processes.
The first four variables are designed
to separate the signal from the \qq\ background.
The last variable provides rejection of the diboson
backgrounds from \Wp\Wm\ and Z$^0$Z$^0$ events surviving
the 6C fit probability requirement (4). Although its use
reduces the efficiency for signals with \mh\ = \mA, the sensitivity
to signals with \mh\ $\neq$ \mA\ is enhanced.
The signal samples used to form reference distributions
for the likelihood are a mixture of samples in the kinematically
accessible region of the (\mh,~\mA) plane with \mh,~\mA~$>$~30~GeV.
The distribution of this likelihood variable for the data,
SM backgrounds, and a representative signal with \mh\ = 30~GeV and
\mA\ = 60~GeV is shown in Figure \ref{fig:taglesshalike}.  The
likelihood variable is required to exceed 0.1 for selected
events. }
\end{itemize}

In Table~\ref{table:taglessha1}
the numbers of events passing the requirements  
after each step, (1) to (5), are given, 
together with the expected SM backgrounds from 4-fermion and 2-fermion
processes. The lowest estimated efficiencies, for \mh\ = 30~GeV and \mA\ = 60~GeV, 
corresponding to the \bb\bb\ final state, and
the number of expected signal events in the 2HDM(II) for the case of 
$\alpha$~= 0, \tanb\ = 1.0, \mh\ = 30~GeV and \mA\ = 60~GeV, are shown in the 
last two columns.

\begin{table}
\begin{center}
\begin{tabular}{|c||r||r||r|r||c||c|}\hline
Cut & Data & Total bkg. & \qq($\gamma$) & 4-fermion & Efficiency [\%] & Signal \\
\hline
\hline
(1) & 1953 & 1885.0  & 537.8 & 1347.2 & 51.2 & 14.9\\
\hline
(2) & 1593 & 1532.0 & 410.1 & 1121.9 & 47.9 & 13.9 \\
\hline
(3) & 1497 & 1446.3  & 376.1 & 1070.2 & 44.3 & 12.9 \\
\hline
(4) & 904  & 895.0 & 329.2 & 565.8 & 40.0 & 11.6 \\
\hline
(5) & 573 & 553.2 & 238.5 & 314.7 & 38.5 & 11.2 \\
\hline
\end{tabular}
\caption{\label{table:taglessha1}
\sl{The \h\A\ channel for the flavour independent analysis: 
the numbers of events selected in the data
at \sqrts\ = 189 GeV,
expected background from SM processes normalised to the data luminosity of 172.1 pb$^{-1}$, 
the minimal detection efficiencies for \mh\ = 30 GeV,
\mA\ = 60 GeV and the expected number of signal events 
within the 2HDM(II) for the case of $\alpha$ = 0,
\tanb\ = 1.0, \mh\ = 30 GeV and \mA\ = 60 GeV after each step of the selection.}}
\end{center}
\end{table}

\begin{figure}[t]
\begin{center}
\epsfig{file=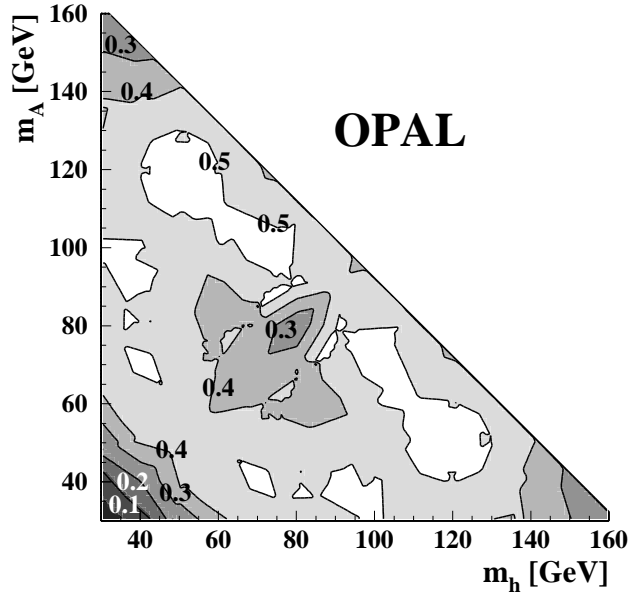,height=9.0cm}
\caption{\label{fig:taglesshaeff}
\sl{Lowest efficiency contours for the flavour independent search for hadronic
\h\A\ events.  The efficiency is taken to be zero for
\mh$\mathrm{+}$\mA$> 189$~GeV.}}
\end{center}
\end{figure}

The selection efficiency is estimated with the Monte Carlo simulation at
a discrete set of reference points in the
(\mh,~\mA) plane.  The efficiency function is interpolated 
by considering the three closest reference points. A plane
in the (\mh,~\mA,~efficiency) space is formed containing those three points,
which allows the efficiency for an arbitrary intermediate (\mh,~\mA) signal 
hypothesis to be computed.  The interpolated efficiency function is shown 
in Figure~\ref{fig:taglesshaeff} for the
final-state flavour assignment with the lowest efficiency.
The efficiency
is low near \mh\ = \mA\ = \mWn\ because of the difficulty in distinguishing
hadronic \h\A\ decays from hadronic \Wp\Wm\ decays.
The veto of \Wp\Wm\  events reduces the efficiency
in the entire (\mh,~\mA) plane because incorrect jet assignments in
\h\A\ events can produce interpretations consistent with the \Wp\Wm\- 
hypothesis.  Conversely, the \Wp\Wm\ veto reduces the background 
of incorrectly-paired \Wp\Wm\ events everywhere in the (\mh,~\mA) plane.
The efficiency for low \mh\ and \mA\ is reduced because of the
requirement that the event should have four distinct jets, which helps 
to reject the \qq\ background.

After the selection, 573 candidates remain in the data, as
compared with the SM expectation of 553.2$\pm$38.2 events.

\hspace{-0.75cm}

\subsection{Discriminant variable}
\label{sect:interpha}

The invariant masses of jet pairs may be 
used to separate possible \h\A\ signals from the \Wp\Wm\ 
and \qq\glgl\ backgrounds.
There are six possible assignments of pairs of jets to 
\h\ and \A\ bosons, and there is no constraint from a known
mass such as that of the \Z\ used in the \ee\ra\ho\Z\ channels.
All six possible interpretations of each selected
event are tested for consistency with a possible signal.
When computing the confidence level for excluding a
model hypothesis, only a single interpretation of each
candidate in the data and Monte Carlo samples may be used.
The choice of jet pairing depends on the \mh\ and \mA\ of the
hypothesised signal.

\begin{figure}[t]
\begin{center}
\epsfig{file=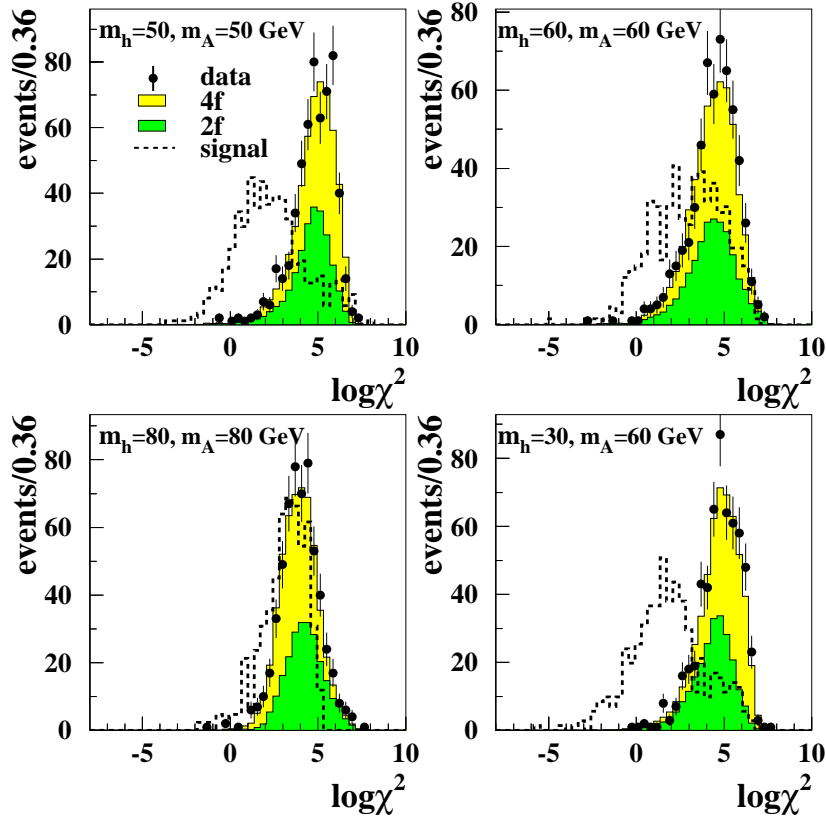,width=12.cm}
\caption{\label{fig:taglesschi2}
\sl{Distribution of $\log\chi^2$ of the mass constraint 
for the flavour independent \ee\ra\h\A\ channel for four
hypotheses of (\mh,~\mA).
The OPAL data are indicated by dots with error bars 
(statistical errors), the 4-fermion backgrounds
by lighter grey histograms and the 2-fermion background
by darker grey histograms.  The \ee\ra\ho\Ao\ra\glgl\glgl\ signals
are shown with arbitrary normalizations by dashed histograms.}}
\end{center}
\end{figure}

For each event passing the selection, a 4C kinematic 
fit constraining energy and momentum conservation is performed. For each of the
assignments of jet pairs to bosons, the reconstructed \mh\ and
\mA\ are computed, along with their covariance matrices.
For each hypothetical \mh\ and \mA\ considered in the limit computation,
each event is assigned the jet pairing with the smallest $\chi^2$ value
resulting from the difference
of the measured and hypothesised \mh\ and \mA, and the error
matrix of the measurement.  The logarithm of the
smallest $\chi^2$ is then used as the discriminating variable when
computing limits because the signal to background ratio depends strongly
on the value of $\log\chi^2$.
Figure~\ref{fig:taglesschi2} shows the distribution
of $\log\chi^2$ for selected data events, the SM expectation,
and the signal for four mass hypotheses.  The signal shown
corresponds to the \ee\ra\ho\Ao\ra\glgl\glgl\ process because of 
its poorer mass resolution
compared with that obtained for final states with quarks.

For \mh\ = \mA\ = \mWn, the separation between the signal and the
background is poor, while for lower values of \mh\ or \mA\ the
separation is better.  The resolution on the reconstructed
sum of the dijet masses is approximately 2.4~GeV, while for
the difference it is approximately 6.2~GeV. The best
sensitivity to the signal is in regions of (\mh,~\mA) with
dijet mass sums different from 2\mWn.  The test mass spacing
is determined by the model scan grid used when 
computing the limits -- no discretization is introduced
within the analysis.  The scan grid used to compute limits
has a finer spacing than the mass resolutions on the candidates.
All candidates are considered at all test mass hypotheses -- they
simply appear at different locations in the $\log\chi^2$ histogram.
The distribution of the
$\chi^2$ variable for the signal and backgrounds changes slowly
with the test mass hypothesis and is interpolated
between Monte Carlo samples generated at different test masses.

Systematic uncertainties have been considered on the signal and background
normalisation and shapes.  The \ee\ra\Wp\Wm\ cross-section
is taken to be uncertain at the level of 2\% from a comparison
of the predictions of the GENTLE and EXCALIBUR calculations.
The selection efficiency for \ee\ra\Wp\Wm\ events is uncertain at the
level of approximately 1\%, from sensitivity to fragmentation modelling
in hadronic W decays and from comparisons of the selection variables
in data and Monte Carlo~\cite{pr183ww}. The background
from \Z/$\gamma\rightarrow$\qq$(\gamma)$~has an 11\% uncertainty~\cite{pr260}, which
includes the uncertainty on the selection efficiency and 
on the four--jet rate in \qq\ events, which is the dominant
contribution.  
The 4-fermion background from two neutral 
vector gauge bosons has been estimated using the grc4f 
Monte Carlo generator for the
central value, and its uncertainty has been
estimated by comparing the results obtained with the grc4f
and EXCALIBUR generators.  
Scaling these uncertainties by
their fractional contributions to the background of this selection
and adding the results in quadrature yields an uncertainty on the
background normalisation of 6.9\%.
Monte Carlo statistics
accounts for only a 1\% relative error on the background.

The uncertainty on the signal efficiencies is dominated by the
flavour dependence, with the highest selection efficiency for 
the \glgl\glgl\ final state. 
The \bb\bb\ and \cc\cc\ final states have very 
similar selection efficiencies.
The lowest signal efficiency at each
mass hypothesis is used in the limit calculations.

A more significant effect on the modelling of the signal is
the uncertainty in the reconstructed mass resolution, as this
affects the shape of the $\log\chi^2$ distribution of the signal
and hence the limits.  Similar
performances are achieved in Monte Carlo simulations of 
the \bb\bb\ and \cc\cc\ final states, but the \glgl\glgl\ final
state has on average a positive shift of one unit of $\log\chi^2$ relative
to the four-quark final states because the resolution is poorer
for reconstructing masses from gluon jets.  The conservative
approach of using the $\log\chi^2$ distribution
of \glgl\glgl\ signal final states
has been adopted when computing the limits.

\section{Model--independent interpretation}
\label{section:modindep}

The results of all the individual search channels
at the studied centre--of--mass energies
are combined statistically to provide 95\% confidence level (CL) limits
in a model--independent interpretation in which 
no assumption is made on the structure of the Higgs sector.
The limits are extracted using the same method applied in previous OPAL publications 
~\cite{smpaper172,mssmpaper172}.

Model--independent \thinspace \thinspace limits \thinspace are \thinspace given \thinspace
for \thinspace the \thinspace \thinspace cross-section \thinspace of \thinspace 
the \thinspace generic \thinspace processes\\ 
\ee\ra~S$^0$\Zo\ and \ee\ra~S$^0$P$^0$, where S$^0$ and 
P$^0$ denote scalar and pseudo-scalar neutral bosons, respectively.\footnote{Throughout this paper numerical mass limits are quoted to 1.0 GeV precision.}
The limits are conveniently expressed in terms of 
scale factors, $s^2$ and $c^2$ ~\cite{pn183}, which relate the cross-sections of these
generic processes to SM cross-sections
(c.f.~Eqs.~(\ref{equation:xsec_zh}),~(\ref{equation:xsec_ah})):
\begin{equation}
\sigma_{\mathrm{SZ}}=s^2~\sigma^{\mathrm{SM}}_{\mathrm{HZ}},
\label{eq:s}
\end{equation}
\begin{equation}
\sigma_{\mathrm{SP}}=c^2~\bar{\lambda}~\sigma^{\mathrm{SM}}_{\mathrm{HZ}}.
\label{eq:c}
\end{equation}

\begin{figure}[t]
\centerline{
\mbox{\epsfig{file=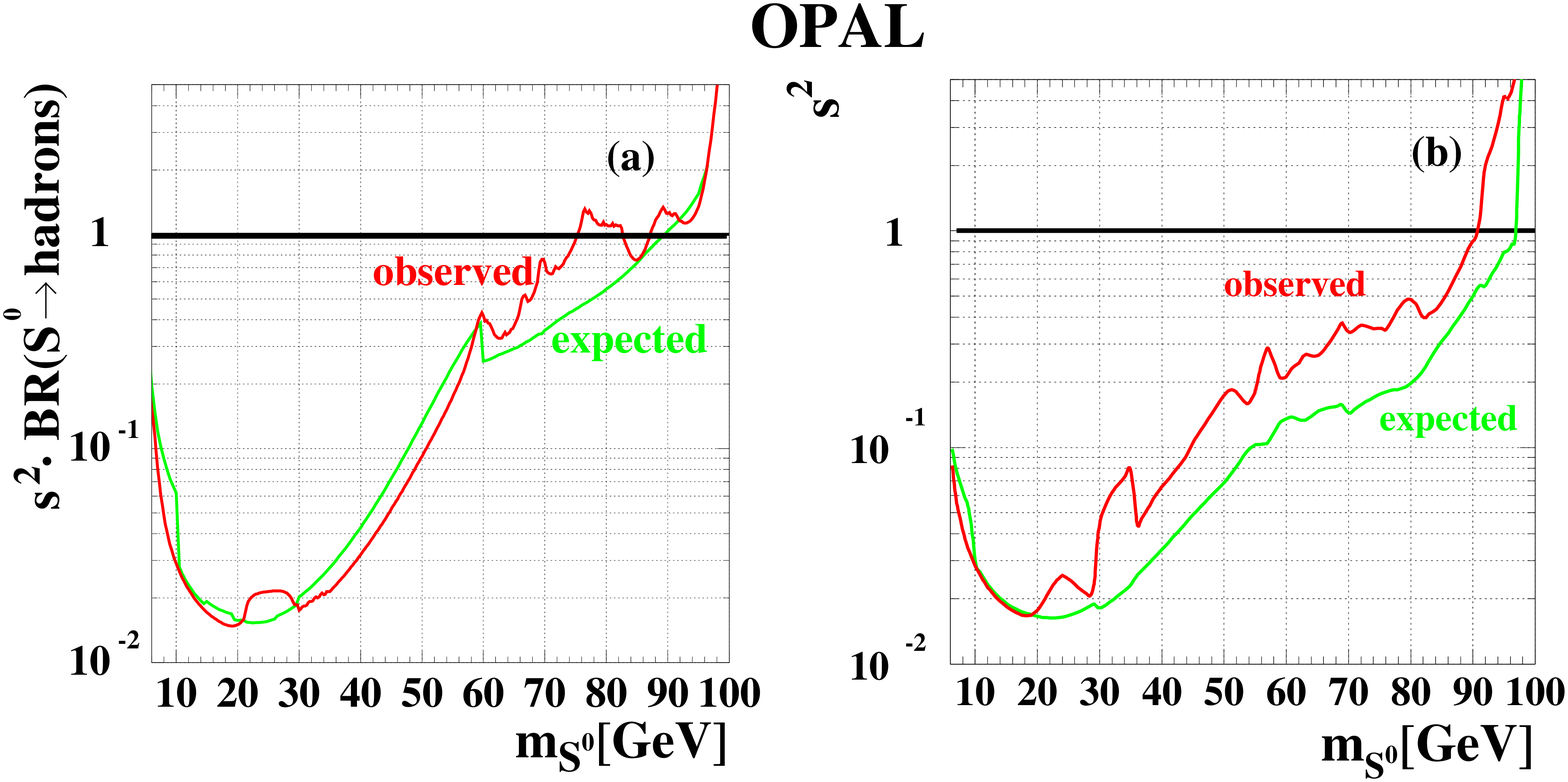,width=18.0cm}}}
\caption[]{\label{modindepZh}\sl 
         Observed and expected upper limits at 95\% CL on (a) 
 $s^2\cdot$BR(\ho\ra\ hadrons) using flavour independent channels, and 
         (b) $s^2$ using all $\mathrm{S^0}$\Zo\ search channels with
 b--tagging and assuming SM Higgs branching ratios for the S$^0$.}
\end{figure}

Figure~\ref{modindepZh} 
shows the 95\% CL upper bound for $s^2$
as a function of the S$^0$ mass, obtained from:
\[s^2=
  \frac{N_{95}^{\mathrm{SZ}}}{
       \sum~(\epsilon~{\cal L}~\sigma^{\mathrm{SM}}_{\mathrm{HZ}})},\]
where $N^{\mathrm{SZ}}_{95}$ is the 95\% CL upper limit
on the number of possible 
signal events in the data, $\epsilon$ is the signal detection efficiency,
${\cal L}$ is the integrated luminosity, and
the sum runs over the different 
centre-of-mass energies of the data and the different channels. 
In Figure~\ref{modindepZh}(a) only the flavour independent channels 
described in Section \ref{sect:zhsearches} and the channels 
analysed at the \Zo\ pole are used to extract a 95\% CL 
upper limit on $s^2\cdot$BR(S$^0$\ra hadrons).  
In Figure~\ref{modindepZh}(b) the SM Higgs branching
ratios for the S$^0$ are assumed and 
search channels with b--tagging are used.
In the region $m_{\mathrm{S}} < 30$ GeV the high energy 
data (LEP2) have little exclusion power while 
for $m_{\mathrm{S}} > 50$ GeV the \Zo\ data (LEP1)
contribute little to the determination 
of the experimental limit.
The $s^2$ limit is calculated only for $m_{\mathrm{S}}$  $\ge 5$~GeV,
since below this mass value the direct search rapidly loses sensitivity 
and the limit is extracted by a different method ~\cite{schaile}, 
which makes use of the electroweak precision measurements of the \Z\ 
width and provides an $s^2$ limit of about $0.5 \times 10^{-2}$.

The limit on $m_{\mathrm{S}}$ for $s^2 =$ 1 assuming SM branching ratios is 91 GeV in complete 
agreement with the result obtained by the SM search ~\cite{pr285} at \sqrts $\approx$ 189 GeV.
A limit of 75 GeV on $m_{\mathrm{S}}$ for $s^2 =$ 1 is obtained when 
assuming a 100\% hadronic branching ratio.
This weaker limit 
is partly 
due to the presence of candidates around $m_{\mathrm{S}}$~$\approx$~80 GeV 
as can be seen from the different behaviour of the observed and expected limit 
in Figure~\ref{modindepZh}(a).

\begin{center}
\begin{figure}[H]
\centerline{\mbox{\epsfig{file=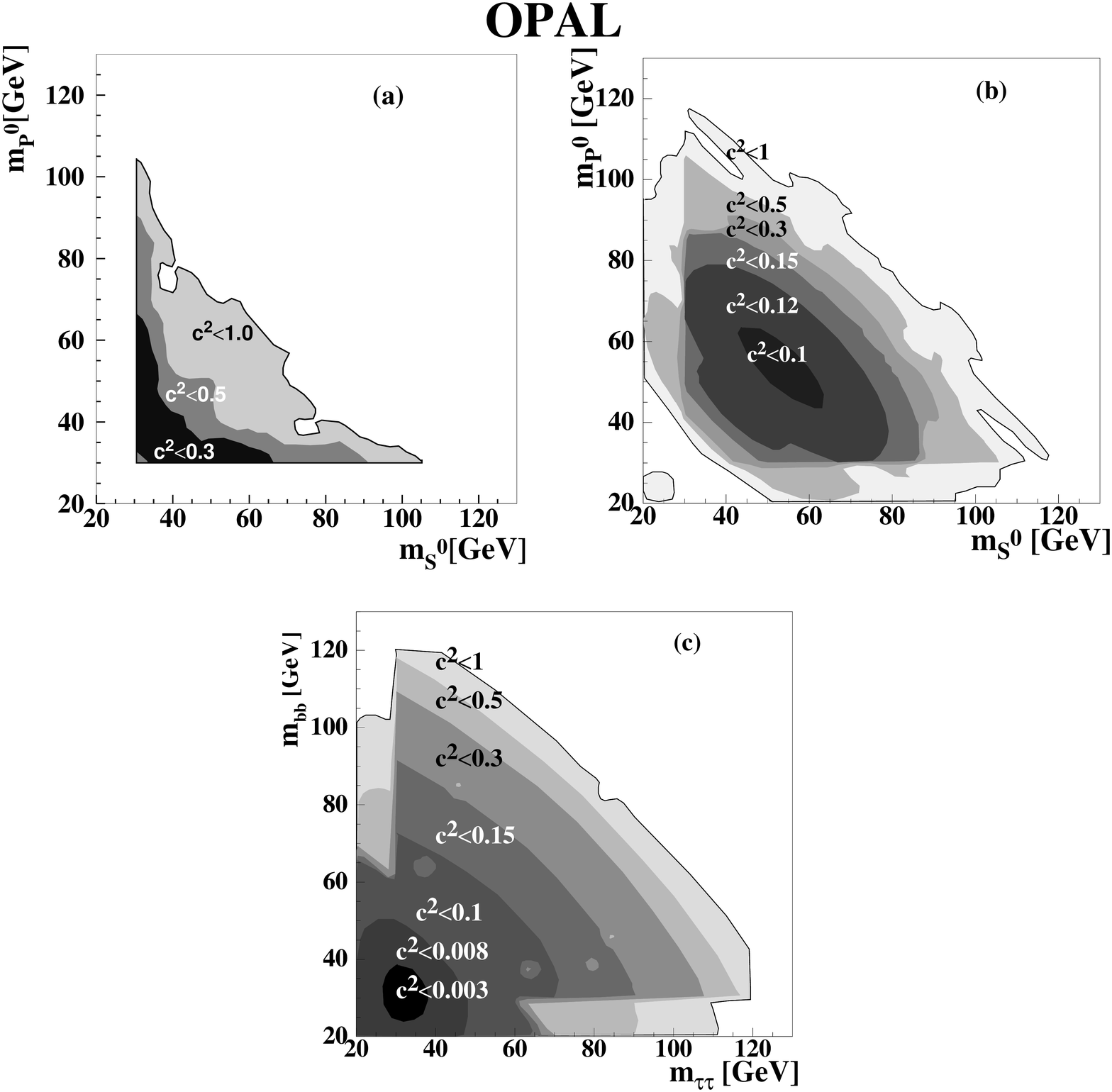,height=16.0cm}}}
\caption[]{\label{modindephA}\sl
         Upper limits at 95\% CL for $c^2$
         (a) for the S$^0$P$^0$\ra\qq\qq, \glgl\qq\ and \glgl\glgl\ 
         search channel assuming the hadronic branching ratio 
         for both S$^0$ and P$^0$ to be 100\%,
         (b) for the S$^0$P$^0$\ra\bb\bb\ search channel assuming the \bb\ branching
         ratio for both S$^0$ and P$^0$ to be 100\%, and 
         (c) for the S$^0$P$^0$\ra\bb\tautau\ search channel assuming a
         100\% branching ratio for this final state.
         The invariant masses of the tau-lepton pair and b--jet pair in (c)
         are denoted by $m_{\tau\tau}$ and $m_{b\bar{b}}$, respectively.
         The iso-contour lines are 
         at values of $c^2 \le 1.0$, 0.5 and 0.3 in (a),
         $c^2 \le 1.0$, 0.5, 0.3, 0.15, 0.12 and 0.1 in (b)
         and $c^2 \le 1.0$, 0.5, 0.3, 0.15, 0.1, 0.008 and 0.003 in (c), 
         respectively.}
\end{figure}
\end{center}

Iso-contours of 95\% CL upper limits for $c^2$ in the S$^0$ and P$^0$ mass plane are shown
for the processes S$^0$P$^0$\ra\qq\qq, \glgl\qq\ and \glgl\glgl\ in Figure~\ref{modindephA}(a), and 
for \ee\ra~S$^0$P$^0$\ra\bb\bb\ and \bb\tautau\ in 
Figures~\ref{modindephA}(b) and (c), respectively, assuming a 100\% branching ratio 
into the specific final states.
The contours are obtained from:
\[c^2=
  \frac{N_{95}^{\mathrm{SP}}}{
       \sum~(\epsilon~{\cal L}~\bar{\lambda}~\sigma^{\mathrm{SM}}_{HZ})},\]
with $N_{95}^{\mathrm{SP}}$ being the 95\% CL upper limit for the number of
signal events in the data.
The results obtained in Figures~\ref{modindephA}(a) and (b) are 
symmetric with respect to interchanging of S$^0$ and P$^0$, while those obtained
for \tautau\bb\ are not. For this reason, the results for
\tautau\bb\ are presented with the 
mass of the particle decaying into \tautau\ along the abscissa and that of the particle
decaying into \bb\ along the ordinate.
The irregularities of the iso-$c^2$ contours are due to the presence
of candidate events.
Along the diagonal, for $c^2 =1$,
a lower bound is extracted using the b--tagging channels
on the masses at $m_{\mathrm{S}} =  m_{\mathrm{P}} \approx m_{\mathrm{\tau^+
\tau^-}} \approx m_{\mathrm{b\bar{b}}}> 78$ 
GeV at 95\% CL. In the hypothesis of S$^0$ P$^0$ decaying to hadrons with a 100\% branching ratio,
a lower bound of $m_{\mathrm{S}} = m_{\mathrm{P}} > 61$ GeV is obtained along the diagonal
for $c^2$ = 1. Note the small region 30 $\le m_{\mathrm{P}},~m_{\mathrm{S}} \le$ 40 GeV in 
Figures~\ref{modindephA}(a) and (b) which is excluded 
by the flavour independent search but not when using only \bb\bb\ channels.

\section{Two Higgs Doublet Model interpretations}
\label{section:limits}

The interpretation of the searches for the neutral Higgs bosons
in the 2HDM(II) is done by scanning the parameter space 
of the model. Every (\mh,~\mA,~\tanb,~$\alpha$) point
determines the production cross--section and 
the branching ratios to different final states. 
An updated version of the HZHA Monte Carlo generator \cite{hzha}
that includes the 2HDM(II) production cross--sections and 
branching ratios for Higgs decays
has been used to scan the parameter space.
This generator includes next-to-next-to-leading order QCD
corrections and next-to-leading order electroweak corrections.
The branching ratios obtained were cross--checked 
with the results of another generator ~\cite{jan}
in which QCD corrections are computed
only up to next-to-leading order.
The comparison showed good agreement between the results of the two
programs.

The results of all the individual search channels\footnote{For the case \ho\Zo\ra\bb\tautau\ and \ho\Zo\ra\tautau\qq\
(tau channel) two different efficiencies are applied, according to the
final state topologies studied.} at the studied centre--of--mass energies
are combined statistically to constrain the 2HDM(II) parameter space.
Although the flavour independent channels supply a unique 
way to investigate parameter space regions where
the branching ratio \ho\ra\bb\ or \Ao\ra\bb\ 
is highly suppressed ({\it{e.g.}}, low $\alpha$
and \tanb\ regions), they have a poor sensitivity with respect to
the b--tagging channels outside these regions.
The use of b--tagging information substantially reduces the background
coming from \WW\ events and improves the sensitivity
to observe Higgs bosons even in regions of the 2HDM(II) parameter 
space where only small branching ratios for \h\ra\bb\ are expected.
The expected confidence 
level is calculated alternatively including only the b--tagged 
or non--b--tagged channels: for each parameter space point,
either the flavour independent or the b--tagging analysis
is then chosen for the extraction of the limits,
depending on which provides the better expected confidence 
level.

The parameter space covered by the present study is:

\begin{itemize}  
\item{$1 \le \mh \le 100$ GeV, in steps of 1 GeV}
\item{$5 \le \mA \le 100$ GeV, in steps of 1 GeV; \\
      $100 \le \mA \le 500$ GeV, in steps of 5 GeV; \\
      $0.5 \le \mA \le 2.0$ TeV, in steps of 0.5 TeV}
\item{$0.4 \le \tanb \le 58.0$, in steps of $1^{\circ}$ in $\beta$}
\item{$\alpha = 0, -\pi/8, -\pi/4, -3\pi/8$ and $-\pi/2$}
\end{itemize}  

The values of $\alpha$ are chosen to extend the analysis 
to the particular cases of maximal and minimal mixing  
in the neutral CP-even sector of the 2HDM(II) ($\alpha = -\pi/4$ and $-\pi/2$, 
respectively)
and of BR(\h\ra\bb) = 0 ($\alpha = 0$).
A more complete picture of the model is obtained
by studying two more intermediate values of $\alpha$.
For \tanb\ $<$ 0.4 radiative corrections become unstable.
Below \mA $\approx$ 5 GeV the direct search 
in the \ee\ra\ho\Ao\ channel cannot be included since 
the detection efficiency vanishes, and the constraint from the total 
\Zo\ width
provides very limited exclusion since the contribution
is too small.
The other two free parameters of the model, 
$m_{\mathrm{H}}$ and $m_{\mathrm{H^{\pm}}}$, are not scanned 
in the present study. They are fixed at values above 
the kinematically accessible region at the present centre--of--mass
energies at LEP2. 

The production of any neutral low mass scalar 
particle in association with the \Zo\ was investigated 
in a previous OPAL publication ~\cite{higgsmall} and, 
for \mh\ $\le$ 9.5 GeV, a mass-dependent upper
limit on the Higgs boson production cross--section
was obtained. This limit translates directly into
an upper limit on the 2HDM(II) production cross--section
for \mh\ below 9.5 GeV. Another powerful experimental constraint
on extensions of the SM is the determination of the total width
of the \Z\ boson at LEP~\cite{schaile}.
Any possible excess width obtained when subtracting
the predicted SM width from the measured $\Gamma_{\mathrm{Z}}$ value can be 
used to place upper limits on the cross--section of 
\Zo\ decays into final states with \ho\ and \Ao\ bosons ~\cite{mssm98}.
An expected increase of the partial width of the \Z\ is 
evaluated for each scanned parameter space point in the 
2HDM(II); if it is found to exceed the experimental limit,
the point is excluded. The two constraints discussed above
are treated together and are referred to as {\it{\Z\ width}}
in the rest of the paper, since for low \mh\ values
most of the excluded regions are obtained from the constraints
derived from $\Gamma_{\mathrm{Z}}$.

The direct searches for the process \ee\ra\h\Z\  (\ee\ra\h\A)
in the \Z\ data contribute 
mainly in the \mh $\le$ 50 GeV (\mh $\le$ 60 GeV) region.
Since the flavour independent \h\Z\ and \h\A\  analyses have been performed
in the mass regions \mh $\ge$ 60 GeV (for the tau and missing energy channel, \mh $\ge$ 30 GeV)
 and  \mh,~\mA $\ge$ 30 GeV, respectively,
 only b-tagging channels using higher energy data are applied below
these masses; however these channels have no detection 
efficiency for \mh $\le$ 30 GeV.
The flavour independent analyses provide exclusion
for the whole \tanb\ range and for the \tanb$<$1 regions
for $\alpha=0$ and  $\alpha=-\pi/8$, respectively.
In Figures \ref{2hdm1}(a--e) the excluded regions in
the (\mh,~\mA) plane are shown for the five chosen values of $\alpha$,
together with the calculated expected exclusion limits.  A particular
(\mh,~\mA, $\alpha$) point is excluded at 95\% CL if it is excluded
for all scanned values of \tanb.  Different domains of \tanb\ are
studied and described below: a) 0.4 $\le$ \tanb $\le$ 58.0 and b) 0.4
$\le$ \tanb\ $\le$ 1.0 or 1.0 $<$ \tanb $\le$ 58.0, for which enlarged
excluded regions are obtained.

\vspace{0.3cm}
\hspace{-.6cm}a) 0.4 $\le$ \tanb $\le$ 58.0 (darker grey area):

\begin{itemize}
\item {The poor sensitivity of the \Z\ channels below \mh\ $\lesssim$ 10 GeV causes
a sharp cut in the exclusion plots at this \mh\ for $\alpha>-\pi/2$.
The exclusion in this region 
is extracted from the total width of the \Z\ boson, as explained 
above. Both the \h\Z\ and \h\A\ production processes contribute to the
natural width of the \Z. 
While an excess,
induced by the \h\Z\ process,
extends the exclusion region to any value of \mA,
the exclusion provided by the \h\A\ process is kinematically limited 
to the region where \mA\ + \mh\ $\le$ \mZ. 
The contribution of the \h\Z\
production cross--section to the \Z\ width 
depends on the argument $(\beta - \alpha)$,
and it becomes large
enough for this process alone to provide exclusion 
in different \tanb\ domains for the 
$\alpha$ values considered.}

\item{ For $\alpha$ = 0 and  $-\pi/8$, 
most of the exclusion is provided by the channels at \sqrts~=~\mZ,
where no b--tagging was applied.
In fact, the flavour independent analyses at \sqrts~ $\approx$ 189 GeV  
have a limited sensitivity because of the presence of the \WW\ background
events. The line at \mh\ $\approx$ 57 GeV in Figure 
\ref{2hdm1}(b) is a result of the \Z\ data kinematic constraint.}
\item { The presence of candidates in the four--jet and \tautau\ 
b-tagging \h\A\ channels
at \sqrts~=~189 GeV at \mh,~\mA\ $\approx$ 80 GeV, due
to the \WW\ background, is clearly reflected in Figure
\ref{2hdm1}(c). The unexcluded region at (\mh,~\mA) $\approx$ (60, 90) GeV in \ref{2hdm1}(d)
is due to the presence of candidates in the four--jet b-tagging \h\A\ channel.}

\item{ The shape of the exclusion plot in Figure \ref{2hdm1}(e)
for \mh\ $<$ 35 GeV is related to the kinematical constraint on the \h\A\
production in the \Z\ data, which for $\alpha$ = $-\pi/2$ and large \tanb\
is the only allowed process, since the \ho\Zo\ production cross--section 
\thinspace vanishes \thinspace when \thinspace $\beta-\alpha\approx\pi$.
\thinspace For \mh\ $>$ 35 \thinspace GeV, \thinspace the \thinspace high \thinspace energy \thinspace data \thinspace
open a \thinspace new 
\begin{figure}[H]
\centerline{ \epsfig{file=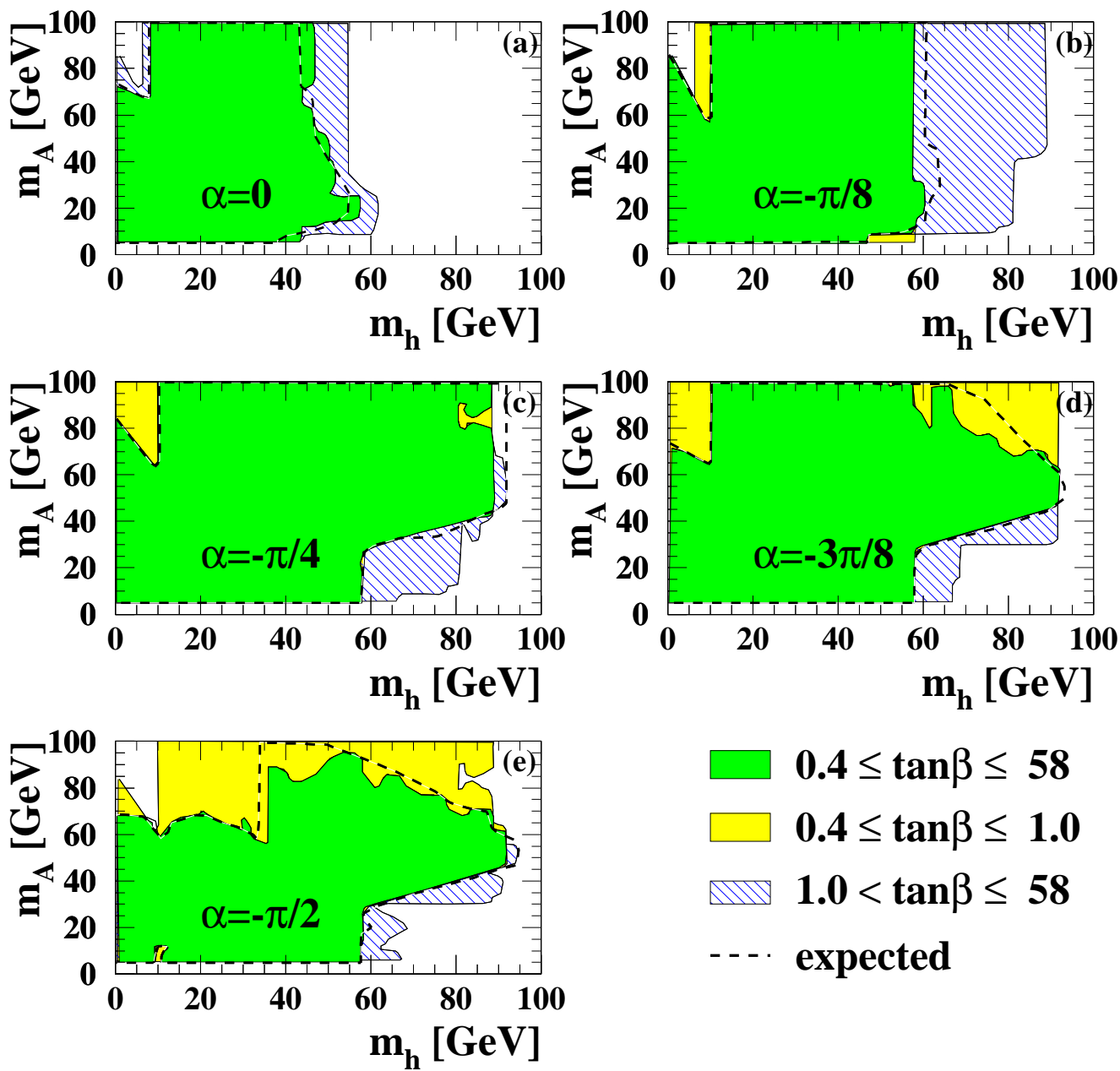,width=20.0cm} }
\caption[] {\label{2hdm1}\sl
Excluded regions in the (\mh,~\mA) plane, (a)--(e), for 
$\alpha=$
0, $-\pi/8$, $-\pi/4$, $-3\pi/8$ and $-\pi/2$, respectively,
together with the expected exclusion limits.
A particular (\mh,~\mA,~$\alpha$) point is excluded at  
95\% CL if it is excluded for all scanned values of \tanb.
Three different domains of \tanb\ are shown:
the darker grey region is excluded for all values 0.4 $\le$ \tanb $\le$ 58.0;
additional enlarged excluded regions are obtained  
constraining 0.4 $\le$ \tanb\ $\le$ 1.0 (lighter grey area) 
or 1.0 $<$ \tanb $\le$ 58.0 (hatched area).
Expected exclusion limits are shown for 0.4 $\le$ \tanb $\le$ 58.0 
(dashed line).}
\end{figure}
kinematic region and are able to exclude large (\mh,~\mA) areas,
as can be seen by the sharp line in Figure \ref{2hdm1}(e).}

\item{ The (\mh,~\mA) points below the semi-diagonal defined by \mh $\ge$ 2\mA,
for which the process \h\ra\A\A\ is kinematically allowed, 
can only be excluded for \tanb\ $>$ 0.5 values by the high energy channels.
In fact, for very low
values of \tanb\ the branching ratio for \A\ra\bb\ vanishes, causing unexcluded regions 
in Figures \ref{2hdm1}(c), \ref{2hdm1}(d) and \ref{2hdm1}(e), which are excluded 
by the \Z\ data flavour independent analyses below \mh\ $\approx$ 60 GeV.}
\end{itemize}
b) 0.4 $\le$ \tanb\ $\le$ 1.0 (lighter grey area) 
and  1.0 $<$ \tanb $\le$ 58.0 (hatched area):
\begin{itemize}
\item { As discussed above, as a consequence of
the variation of the \h\Z\ 
production cross--section with \tanb\ in the \mh\ $<$ 10 GeV region,   
for $\alpha ~=~ -\pi/8$ in Figure \ref{2hdm1}(b),  
the \mh~$>$ 7 GeV region is excluded for all values of \mA\ in the
\tanb\ $\le$ 1.0 domain. 
For $\alpha$ = $-\pi/4$ and $-3/8\pi$, 
the \mh~$<$ 10 GeV region is excluded for all values of \mA\ only in
the \tanb\ $\le$ 1.0 domain.}
\item{ At $\alpha=0$ and $\alpha=-\pi/8$ and small values of \tanb\ 
the production cross--section for the process \ee\ra\h\Z\
is highly suppressed. For \mh~$>$ 40 GeV, constraining
\tanb~$>$ 1.0,  larger excluded regions are obtained, 
as can be seen in Figures \ref{2hdm1}(a) and (b) (hatched areas).}

\item {The presence of candidates in the four--jet and \tautau\ b-tagging \h\A\ channels
at \sqrts = 189 GeV at \mh,~\mA\ $\approx$ 80 GeV, corresponding
to the \WW\ background, is clearly reflected in Figures 
\ref{2hdm1}(c), (d) and (e).
For $\alpha$ = $-\pi/4$ and $\alpha$ = $-3\pi/8$
this region is unexcluded even for \tanb\ $>1.0$, while for \tanb\ $\le$ 1.0 it is excluded 
due to a large expected production cross--section.
For $\alpha$ = $-\pi/2$, the production cross-section becomes small and
this domain is unexcluded for \tanb\ $\le$ 1.0, as can be seen in Figure \ref{2hdm1}(e).}
\end{itemize}

\begin{figure}[t]
\centerline{ \epsfig{file=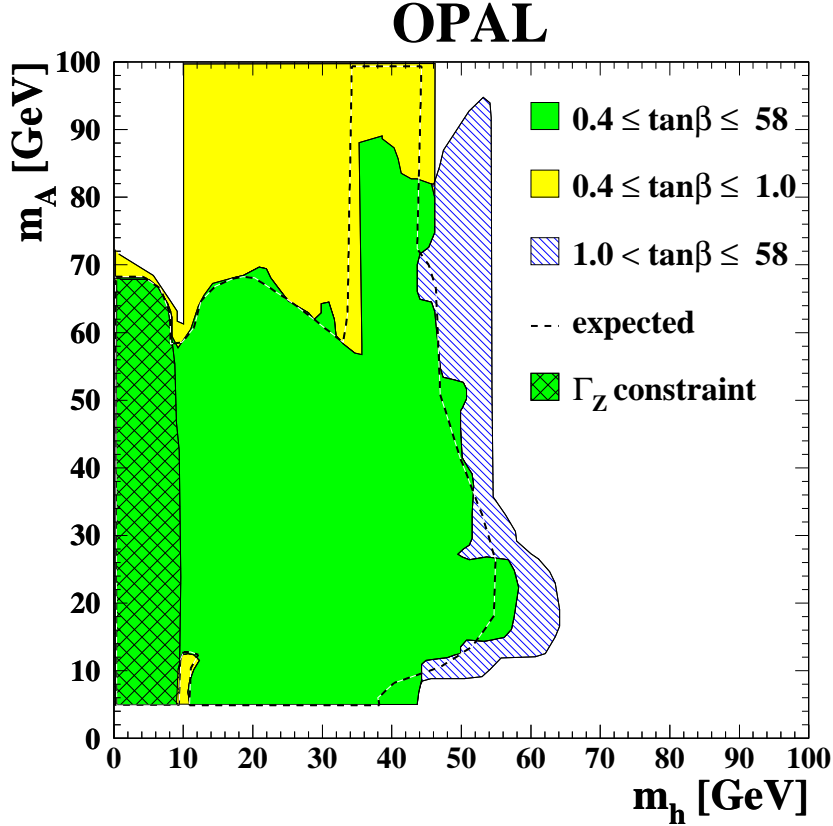,width=12.0cm} }
\caption[]{\label{massmin}\sl
Excluded (\mA,~\mh) region independent of $\alpha$, together with the expected exclusion limit.
A particular (\mA, \mh) point is excluded at  
95\% CL if it is excluded for 0.4 $\le$ \tanb\ $\le$ 58.0 (darker grey region),
0.4 $\le$ \tanb\ $\le$ 1.0 (lighter grey region) and
1.0 $<$ \tanb\ $\le$ 58.0 (hatched region) for any $\alpha$. The cross--hatched
region is excluded using constraints from  $\Gamma_{\mathrm{Z}}$ only.
Expected exclusion limits are shown as a dashed line.}
\end{figure}

In Figure \ref{massmin} the excluded regions in the (\mh,~\mA) plane
independent of $\alpha$ are given together with the calculated expected 
exclusion limits.
A particular (\mh, \mA ) point is excluded at  
95\% CL if it is excluded for all scanned values of \tanb\ and $\alpha$.
Different domains of \tanb\ are
shown: 0.4 $\le$ \tanb\ $\le$ 58.0 (darker grey area), 0.4
$\le$ \tanb\ $\le$ 1.0 (lighter grey area) and 
1.0  $<$ \tanb\ $\le$ 58.0 (hatched area), for which enlarged
excluded regions are obtained.
The rectangular region 1 $\lesssim$ \mh\ $\lesssim$ 44 GeV for
12 $\lesssim$ \mA\ $\lesssim$ 56 GeV is excluded at 95\% CL
independent of $\alpha$ and \tanb.
The cross hatched region shows the exclusion provided by the constraints on 
the width of the \Zo\ common to all the scanned values of $\alpha$ and \tanb.

\begin{figure}[H]
\centerline{ \epsfig{file=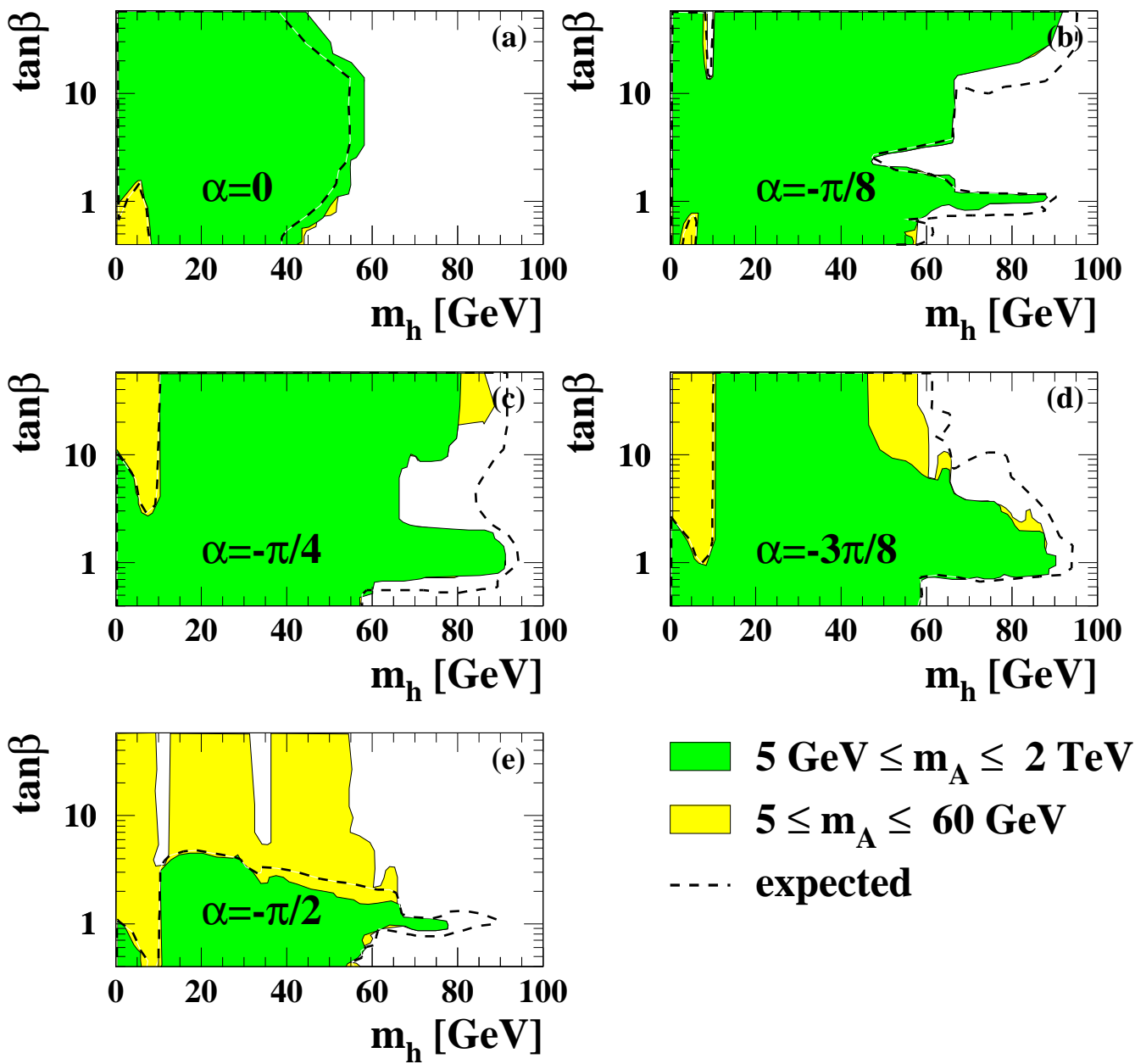,width=20.0cm} }
\caption[]{\label{2hdm2}\sl
Excluded regions in the (\tanb, \mh) plane,
(a)--(e), for $\alpha$=
0, $-\pi/8$, $-\pi/4$, $-3\pi/8$ and $-\pi/2$, respectively,
together with the expected exclusion limits.
A particular (\mh, \tanb, $\alpha$) point is excluded at  
95\% CL if it is excluded for all scanned values of \mA.
The two regions shown correspond to the whole domain 5 GeV $\le$ \mA $\le$ 2 TeV (darker grey area)
and a restricted domain for which 5 $\le$ \mA\ $\le$ 60 GeV (lighter grey area). 
The exclusion regions for \mA\ $\le$ 60 GeV entirely contain
the  5 GeV $\le$ \mA $\le$ 2 TeV excluded areas.
Expected exclusion limits are shown for 5 GeV $\le$ \mA $\le$ 2 TeV
(dashed line).
}
\end{figure}

In Figures \ref{2hdm2}(a--e) the excluded regions in the (\tanb, \mh) plane are shown
for the five chosen values of $\alpha$,
together with the calculated expected exclusion limits.
A particular (\mh, \tanb, $\alpha$) point is excluded at  
95\% CL if it is excluded for all scanned values of \mA.
There are two regions shown, the whole domain 5 GeV $\le$ \mA $\le$ 2 TeV (darker grey area)
and a restricted domain for which 5 $\le$ \mA\ $\le$ 60 GeV (lighter grey area). The exclusion contours 
for \mA\ $\le$ 60 GeV are larger for all $\alpha$ values, and entirely contain
the  5 GeV $\le$ \mA $\le$ 2 TeV excluded areas.

In Figures \ref{2hdm2}(a) and (b), the unexcluded regions at low \tanb\ and \mh\ $\le$ 10 GeV
reflect the behaviour in Figures \ref{2hdm1}(a) and (b) in the (\mh,~\mA) 
plane for the same values of $\alpha$, for \mh\ $\le$ 10 GeV and \mA\ $>$ 60 GeV.
Similarly, the remaining three values of $\alpha$ show a complementary behaviour at large \tanb,
reflecting the corresponding exclusion regions in the (\mh,~\mA) plane, Figures
\ref{2hdm1}(c), (d) and (e). 
In the \mA\ $\le$ 60 GeV contours 
the same regions are obviously excluded.
The \h\Z\ production cross--section contribution to the \Z\ width is not large enough 
to exclude the small region \tanb $>$ 10, 7 $\le$ \mh\ $\le$ 10 GeV in Figure \ref{2hdm2}(b). In Figure \ref{2hdm2}(e) the two unexcluded regions in the lighter grey area (5 $\le$ \mA\ $\le$ 60 GeV)
at \mh\ $\approx$ 10 and 35 GeV are projections of the two 
regions unexcluded at the same \mh\ values in Figure \ref{2hdm1}(e) (darker grey area), for 
\mA\ $<$ 10 GeV and \mA\ $<$ 60 GeV, respectively. 

\begin{figure}[H]
\centerline{ \epsfig{file=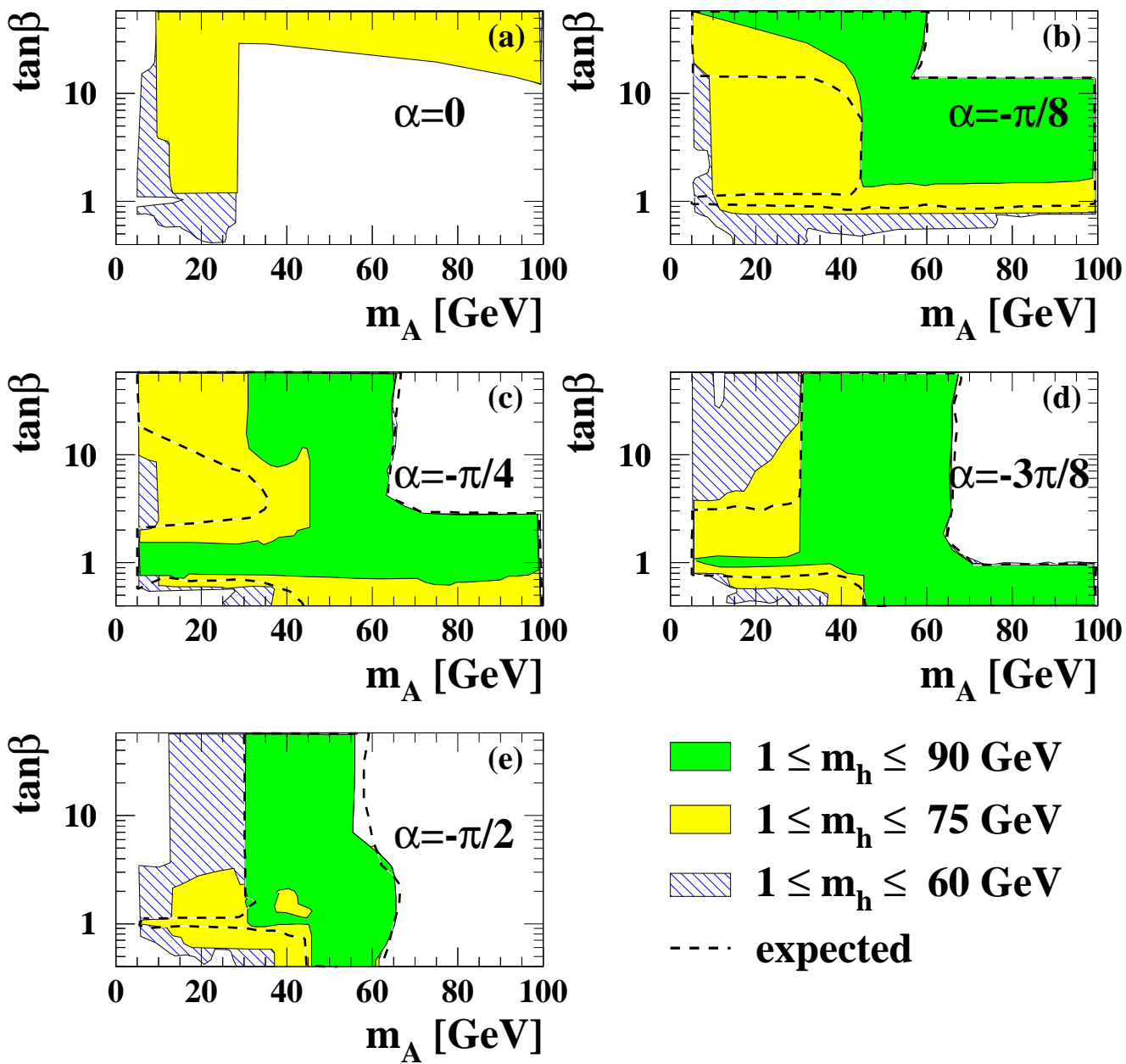,width=20.0cm} }
\caption[]{\label{2hdm3}\sl
Excluded regions in the (\mA, \tanb)
plane, (a)--(e), for $\alpha$=
0, $-\pi/8$, $-\pi/4$, $-3\pi/8$ and $-\pi/2$ respectively,
together with the calculated expected exclusion limits. 
A particular (\mA, \tanb, $\alpha$) point is excluded at  
95\% CL if it is excluded for all scanned values of \mh.
The three contours correspond to
1 $\le$ \mh\ $\le$ 90 GeV (darker grey area),
1 $\le$ \mh\ $\le$ 75 GeV (lighter grey area) and 
1 $\le$ \mh\ $\le$ 60 GeV (hatched area). 
Expected exclusion limits are shown for 1 $\le$ \mh\ $\le$ 90 GeV
(dashed line).
}
\end{figure}

In Figure \ref{2hdm3} the excluded regions in the (\mA, \tanb)
plane are shown for different values of $\alpha$,
together with the expected exclusion limits. 
A particular (\mA, \tanb, $\alpha$) point is excluded at  
95\% CL if it is excluded for all scanned values of \mh.
There are three regions shown, corresponding to different 
\mh\ domains that are subsets of one another, namely: 
1 $\le$ \mh\ $\le$ 90 GeV (darker grey area),
1 $\le$ \mh\ $\le$ 75 GeV (lighter grey area) and 
1 $\le$ \mh\ $\le$ 60 GeV (hatched area). 
The lower the \mh\ upper value analysed, 
the larger the 
excluded (\mA, \tanb) region.
The unexcluded regions for \mA\ $>$ 60 GeV in Figures \ref{2hdm3}(b--e) 
gradually increase with decreasing $\alpha$. 
There is an exact correspondence between 
the largest excluded
\tanb\ value and the excluded regions at \mh\ $<$ 10 GeV in the (\mh,~\tanb)

\newpage

\newpage
\newpage

\hspace{-0.75cm}
projections in Figures \ref{2hdm2}(b--e).
The line at \mA\ $\approx$ 30 GeV for \tanb\ $>$ 1.0
in the exclusion region for \mh\ $<$ 90 GeV in Figures \ref{2hdm3}(d)
and (e) corresponds to the horizontal line at \mA\ = 30 GeV in the 
\tanb\ $>$ 1.0 contour in Figures \ref{2hdm1}(d) and (e).  
The small island at 35 $<$ \mA\ $<$ 39 GeV and 1.0 $<$ \tanb\ $<$ 2.0
in Figure \ref{2hdm3}(e) is the reflection of a candidate in the 
\h\A\ra\tautau\bb\ channel at \sqrts = 183 GeV with \mA\ $\approx$ 37 GeV
and \mh\ $\approx$ 80 GeV.
\section{Conclusions} 
\label{section:conclusion}

A general analysis of the 2HDM(II) 
with no CP--violation and no extra particles 
besides those of the SM and the five Higgs bosons
has been performed for the first time.
Large areas of the parameter 
space of the model have been scanned. 
In the scanning procedure 
the dependence of the 
production cross--sections and branching ratios
on the angles $\alpha$ and $\beta$,
calculated with next-to-next-to-leading order QCD
corrections and next-to-leading order electroweak corrections, has been considered. 

In addition to the standard OPAL b--tagging analyses, 
new flavour independent channels for both the 
Higgs--strahlung process, \ee\ra\ho\Zo, 
and the pair--production process, \ee\ra\ho\Ao,
have been analysed, providing access to regions of parameter 
space in the 2HDM(II) where \ho\ and \Ao\ 
are expected to decay predominantly into up--type light quarks
and gluons ({\it{e.g.}} $\alpha \approx 0$). 

OPAL data collected at \sqrts\ $\approx$ \mZ, 183 and 189 GeV have
been interpreted both in the context of the 2HDM(II) and in a 
model--independent approach where both 
SM branching ratios and 100\% hadronic branching ratios 
were assumed. 

The 2HDM(II) parameter space scan, for 1 $\le$ \mh\ $\le$ 100 GeV,
5 GeV $\le$ \mA\ $\le$ 2 TeV, $-\pi/2\le\alpha\le0$ and $0.4\le\tanb\le58.0$,
leads to large regions
excluded at the 95\% CL in the (\mh, \mA) plane
as well as in the (\mh, \tanb) and (\mA, \tanb)
projections. The region 1 $\lesssim$ \mh\ $\lesssim$ 44 GeV and
12 $\lesssim$ \mA\ $\lesssim$ 56 GeV is excluded at 95\% CL
independent of $\alpha$ and \tanb\ within the 
scanned parameter space.

In the model--independent approach for \ee\ra~S$^0$\Zo, lower bounds at
95\% CL are obtained for $s^2 =$ 1 of 91 GeV assuming SM branching ratios
and 75 GeV assuming 100\% hadronic branching ratios. 
In the case of the generic processes S$^0$P$^0$ $\rightarrow$ \bb\bb\
and S$^0$P$^0~\rightarrow$~\bb~$\tau ^+ \tau ^-$, 
a lower bound at 95\% CL of $m_{\mathrm{S}} = m_{\mathrm{P}} > 78$ GeV
is extracted along the diagonal for $c^2 = 1 $ assuming 
100\% branching ratios for the individual final states, while
assuming the S$^0$ P$^0$ hadronic branching ratios to be 100\%
gives $m_{\mathrm{S}} = m_{\mathrm{P}} > 61$ GeV.

\vspace{1.5cm}
\hspace{-.75cm}
{\large{\bf{Acknowledgements:}}}\\

We particularly wish to thank the SL Division for the efficient operation
of the LEP accelerator at all energies
 and for their continuing close cooperation with
our experimental group.  We thank our colleagues from CEA, DAPNIA/SPP,
CE-Saclay for their efforts over the years on the time-of-flight and trigger
systems which we continue to use.  In addition to the support staff at our own
institutions we are pleased to acknowledge the  \\
Department of Energy, USA, \\
National Science Foundation, USA, \\
Particle Physics and Astronomy Research Council, UK, \\
Natural Sciences and Engineering Research Council, Canada, \\
Israel Science Foundation, administered by the Israel
Academy of Science and Humanities, \\
Minerva Gesellschaft, \\
Benoziyo Center for High Energy Physics,\\
Japanese Ministry of Education, Science and Culture (the
Monbusho) and a grant under the Monbusho International
Science Research Program,\\
Japanese Society for the Promotion of Science (JSPS),\\
German Israeli Bi-national Science Foundation (GIF), \\
Bundesministerium f\"ur Bildung und Forschung, Germany, \\
National Research Council of Canada, \\
Research Corporation, USA,\\
Hungarian Foundation for Scientific Research, OTKA T-029328, 
T023793 and OTKA F-023259.\\

\bibliographystyle{phaip}
\bibliography{higgs}
\end{document}